\documentclass[fleqn, usenatbib, twocolumn, twocolappendix]{openjournal}

\usepackage{lipsum}

\usepackage{xcolor}
\usepackage{textgreek}
\usepackage[utf8]{inputenc}
\usepackage[english]{babel}

\usepackage{hyperref}
\hypersetup{
    unicode, 
    colorlinks=true,
    linkcolor=linkcolor,
    citecolor=linkcolor,
    filecolor=linkcolor,
    urlcolor=linkcolor,
}
\usepackage{color,colortbl}
\definecolor{linkcolor}{rgb}{0.0,0.3,0.5}
\usepackage{tensind}
\tensordelimiter{?}
\DeclareGraphicsExtensions{.bmp,.png,.jpg,.pdf}
\usepackage{verbatim}
\usepackage[normalem]{ulem}
\usepackage{orcidlink}
\usepackage{soul}

\graphicspath{ {./figs/} }

\usepackage{graphicx}	
\usepackage{amsmath}	
\usepackage{amssymb}	
\usepackage[normalem]{ulem} 
\usepackage{hyperref}
\usepackage{tabularx}

\usepackage{lipsum} 
\usepackage{multirow}
\DeclareUnicodeCharacter{22C6}{$\star$}

\newcommand{\eagle}{\mbox{\sc{Eagle}}}

\newcommand{\flares}{\mbox{\sc Flares}}

\newcommand{\jwst}{\mbox{\it JWST}}

\begin{document}
\title[FLARES XXI: UV indices]{First Light And Reionization Epoch Simulations (FLARES) XXI: \\
The UV Indices of Galaxies in the Early Universe}


\author{
Connor Sant Fournier\orcidlink{0000-0000-0000-0000}\textsuperscript{1,⋆},
Stephen M. Wilkins\orcidlink{0000-0003-3903-6935}\textsuperscript{1,3,†},
Joseph Caruana\orcidlink{0000-0000-0000-0000}\textsuperscript{2,1},
Kristian Zarb Adami\orcidlink{0000-0000-0000-0000}\textsuperscript{2,1}, 
Conor M. Byrne\orcidlink{0000-0002-6853-4055}\textsuperscript{4},
William J. Roper\orcidlink{0000-0002-3257-8806}\textsuperscript{3},
Jack C. Turner\orcidlink{0000-0001-6247-041X}\textsuperscript{3}, and 
Aswin P. Vijayan\orcidlink{0000-0002-1905-4194}\textsuperscript{3},
}

\affiliation{\textsuperscript{1}Institute of Space Sciences and Astronomy, University of Malta, Msida MSD 2080, Malta}
\affiliation{\textsuperscript{2}Department of Physics, Faculty of Science, University of Malta, Msida MSD 2080, Malta}
\affiliation{\textsuperscript{3}Astronomy Centre, University of Sussex, Falmer, Brighton BN1 9QH, UK}
\affiliation{\textsuperscript{4}Department of Physics, University of Warwick, Gibbet Hill Road, Coventry CV4 7AL, UK}

\email{connor.sant.17@um.edu.mt} 
\email{s.wilkins@sussex.ac.uk}

\begin{abstract}
UV absorption line indices trace chemical enrichment and star formation histories in high-redshift galaxies, yet their reliability as quantitative stellar metallicity (Z$_\star$) diagnostics remains uncertain. In this work, we combine synthetic spectral modelling with cosmological simulations to assess the behaviour of rest-frame UV indices in the early Universe. Using the forward-modelling package \texttt{Synthesizer}, we compute equivalent widths for UV indices based on BPASS stellar population synthesis models and examine their sensitivity to metallicity, star formation history, nebular emission, and model assumptions. We first investigate idealised stellar populations to establish the metallicity dependence of each index. Most indices show increasing equivalent width with metallicity, although the strength and linearity of this relation varies between features. The 1719~\text{\AA} index exhibits one of the most stable correlations with stellar metallicity, while the 1460~\text{\AA} index shows stronger sensitivity to nebular emission, bursty star formation, and model-dependent effects at high metallicity. We then apply these models to galaxies from the First Light and Reionization Epoch Simulations (\flares), which provide realistic star formation and chemical enrichment histories. The simulated galaxy populations reproduce a stellar mass--metallicity relation and allow the UV indices to be tested in composite spectra. Although scatter increases due to population mixing and stochastic enrichment, the overall metallicity trends remain largely preserved. These predictions provide a theoretical framework for interpreting rest-frame UV spectra from \textit{JWST} and future surveys, supporting the use of UV absorption indices as complementary tracers of stellar metallicity in early galaxies.
\end{abstract}

\keywords{methods: numerical -- galaxies: formation -- galaxies: evolution -- galaxies: high-redshift -- galaxies: extinction -- infrared: galaxies}



\section{Introduction} \label{sec:intro}
A central goal of extragalactic astrophysics is to trace the build-up of elements across cosmic time. Measurements of chemical enrichment provide important constraints on stellar evolution, feedback processes, and galaxy formation models.

Since its launch, the James Webb Space Telescope \citep[\textit{JWST,}][]{gardner2023james} has transformed the study of high-redshift galaxies \citep{Bunker23, bagley2024next, holwerda2024cosmic}. Operating in the infrared, \textit{JWST} enables spectroscopic and photometric observations of galaxies in the early Universe, including systems observed less than 400 million years after the Big Bang. These discoveries challenge existing models of early galaxy formation by revealing unexpectedly luminous and rapidly assembled galaxies at very high redshift \citep[][]{boylan2023stress, lovell2023extreme}.

The sensitivity and angular resolution of \textit{JWST} enable detailed studies of the internal structure, star formation activity, and chemical composition of early galaxies. These observations indicate that some high-redshift systems already exhibit significant chemical enrichment and well-defined morphologies, suggesting rapid stellar mass assembly and structural evolution at early cosmic times \citep[e.g.][]{ferreira2022panic, ferreira2023jwst, kartaltepe2023ceers}.

In addition, \textit{JWST} provides new constraints on galaxies during the epoch of reionization, when the first stars and galaxies ionised the intergalactic medium. Observations of galaxies in this period enable measurements of the chemical composition and physical conditions of the ionised gas through nebular emission lines. Strong oxygen transitions, such as [O~\textsc{iii}]~($\lambda$ 4959,5007) and [O~\textsc{ii}]~($\lambda$ 3726,3729), are particularly valuable diagnostics of gas-phase metallicity and ionisation conditions \citep{shapley2023jwst}.

However, the chemical composition of a galaxy is not described by a single metallicity. Instead, metals are distributed across different components, including stars, cold star-forming gas, and hot gas, each of which traces a distinct enrichment history and requires different observational or modelling diagnostics. In this work, we focus on stellar metallicity.

Stellar metallicity can be constrained using several rest-frame UV diagnostics, including absorption features \citep{Rix04,Sommariva2012stellar,Calabro21,Byrne21}, the continuum slope \citep{VandelsCullen}, spectral breaks \citep{morales2025testing}, and flux ratios \citep{senchyna2017ultraviolet}. Here, we investigate UV continuum indices.

Specific rest-frame UV continuum indices have been identified as tracers of stellar metallicity. These features, spanning $1370 \leq \lambda/\text{\AA} \leq 1853$, were initially defined by \citet{fanelli1992spectral} and later revised by \citet{Calabro21}. Their equivalent widths encode the strengths of metal absorption features in the integrated stellar spectrum, providing a means of tracing the chemical enrichment of stellar populations in high-redshift galaxies.

The UV indices compiled by \citet{fanelli1992spectral} were subsequently examined by \citet{Rix04} and \citet{leitherer2011ultraviolet}, who focussed on absorption features near 1370~\text{\AA}, 1425~\text{\AA}, and 1978~\text{\AA}. These studies show that rest-frame UV absorption indices provide a promising route for probing the stellar populations of young, high-redshift galaxies. In particular, indices with strong metallicity sensitivity offer a means of constraining the chemical composition of unresolved stellar populations.

However, the reliability of UV indices as metallicity diagnostics remains uncertain, primarily because several features are affected by absorption from the interstellar medium (ISM) \citep{Rix04, leitherer2011ultraviolet}. Earlier studies are also limited by small sample sizes, modest spectral resolution, and restricted redshift coverage. The 1978~\text{\AA} index, for example, is sensitive to variations in the initial mass function (IMF) \citep{Sommariva2012stellar}. By contrast, the 1460~\text{\AA}, 1501~\text{\AA}, and 1533~\text{\AA} indices, associated with absorption from species such as Ni~\textsc{II}, S~\textsc{V}, and Si~\textsc{II} in the photospheres of young, hot stars, are less sensitive to stellar age and IMF variations, making them more robust tracers of stellar metallicity \citep{Calabro21}.

Recent advances in high-redshift spectroscopy, particularly with \textit{JWST}/NIRSpec, now enable rest-frame UV diagnostics to be tested using spectra of distant, young galaxies. Early \textit{JWST} observations have spectroscopically confirmed galaxies deep into the reionization era, while also revealing complex rest-frame UV and optical properties, including strong nebular emission, blue UV continua, high ionisation conditions, and rapid chemical enrichment \citep{curti2023jades, Bunker23, Carniani2024, saxena2026hitting}. Large spectroscopic samples from surveys such as the JWST Advanced Deep Extragalactic Survey (JADES) now provide the first opportunity to examine UV absorption and emission features across statistically meaningful populations of galaxies at $z \gtrsim 5$ \citep{Bunker23,roberts2024between}.

These observations motivate a renewed assessment of UV absorption indices as stellar metallicity tracers. While nebular emission lines provide valuable constraints on the ionised gas, stellar metallicity remains more difficult to measure directly at high redshift. Rest-frame UV absorption features offer a complementary route, as they are imprinted by the integrated massive stellar population and are accessible with \textit{JWST}/NIRSpec for galaxies in the early Universe. Revisiting the UV indices in this context is therefore essential for determining which features remain reliable in young, chemically evolving systems, and for establishing how they can be used alongside nebular diagnostics to interpret the growing body of \textit{JWST} spectra.

In this paper, we re-evaluate UV indices from the recent literature using the synthetic observations package \texttt{Synthesizer} \citep{lovell2025synthesizer, roper2025synthesizer} and the First Light and Reionization Epoch Simulations \citep[\flares,][]{FLARES-I,FLARES-II}.

Building on previous investigations \citep{Rix04, Sommariva2012stellar, Calabro21}, we use the Binary Population and Spectral Synthesis model (BPASS) \citep{BPASS2.2.1} to generate theoretical simple stellar populations. We first examine how the equivalent widths of UV absorption indices respond to variations in intrinsic physical parameters at fixed metallicity. Specifically, we investigate the effects of star formation history (SFH), initial mass function (IMF), nebular emission, and spectral grid resolution. We also compare the sampling of the BPASS grids with the resolving power of \textit{JWST}/NIRSpec in order to assess how model resolution affects the measurement and interpretation of UV indices.

We then extend the analysis to the \flares\ suite of cosmological simulations, which provides galaxies with diverse star formation and chemical enrichment histories. Using these simulated galaxies, we investigate how UV index strengths vary across realistic composite stellar populations. We also examine the relationship between stellar mass, star formation, and different definitions of stellar metallicity, including mass-weighted, luminosity-weighted, and young-star-weighted metallicities. This provides the context required to interpret how the complexity of simulated galaxy populations modifies the behaviour of UV indices relative to idealised simple stellar population models.

The paper is structured as follows. Section \ref{sec:theory} examines the dependence of UV index equivalent widths on metallicity, star formation history, IMF, nebular emission, and spectral resolution using \texttt{Synthesizer}. Section \ref{sec:predictions} presents predictions from \flares, including the mass--metallicity relation and the connection between UV indices and galaxy physical properties. Finally, Section \ref{sec:conc} summarises the main findings and conclusions.

\section{Theoretical Background}\label{sec:theory}
\begin{figure*}
    \centering
    \includegraphics[width=\linewidth]{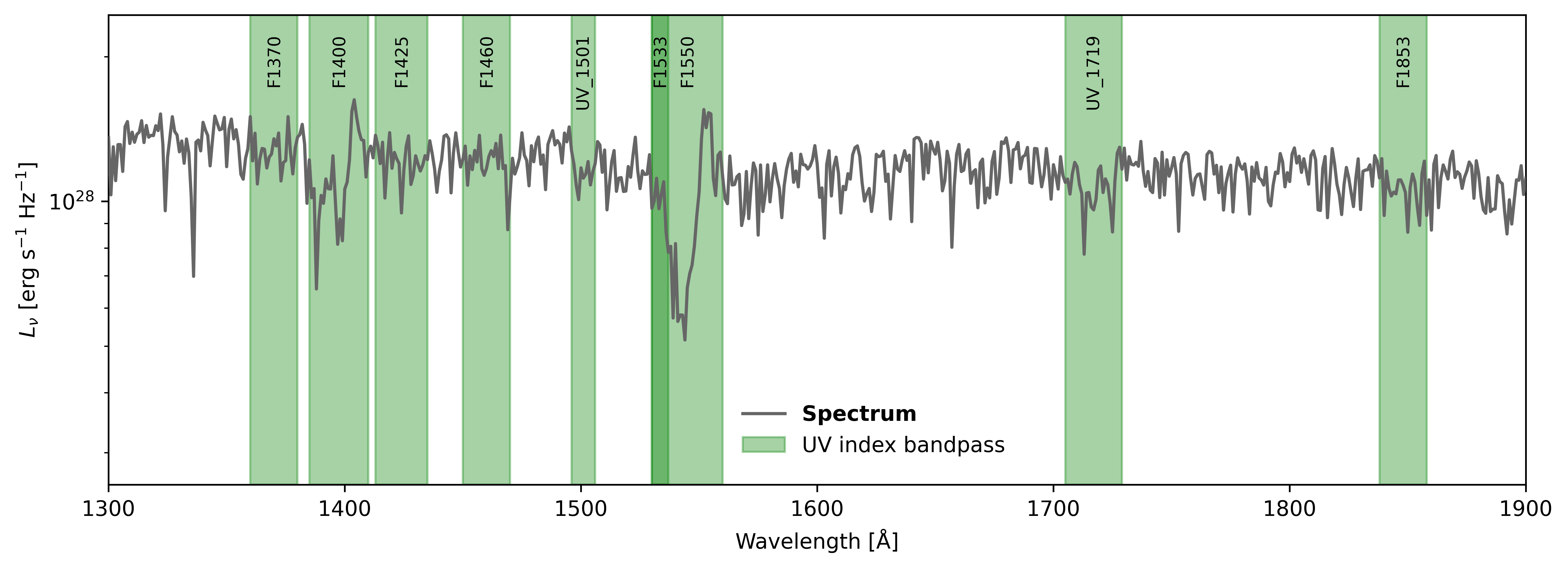}
    \caption{Illustration of a modelled stellar population spectrum at metallicity $Z = 0.04$, generated assuming constant star formation over a stellar population age of 100 Myr. Each UV index bandpass is denoted by a green band.}
    \label{fig:Figure Pseudo}
\end{figure*}

In this section, we use simplified star formation and chemical enrichment histories to examine how each UV index varies with stellar metallicity. We then assess the sensitivity of these indices to star formation history, initial mass function, nebular emission, and spectral grid resolution. This analysis is carried out using the \texttt{Synthesizer} synthetic observations package \citep{lovell2025synthesizer, roper2025synthesizer}.

\subsection{The UV indices}
Continuum UV indices centred near 1400~\text{\AA} and 1550~\text{\AA} are commonly used to investigate metal enrichment in high-redshift galaxies, including systems at z $\sim$ 5 \citep{leitherer2011ultraviolet, faisst2016rest}. The 1550~\text{\AA} index is often treated as a metallicity-sensitive feature because of its association with C~\textsc{iv} absorption, which can exhibit a P-Cygni profile produced by stellar winds \citep{castor1979atlas}. The 1370~\text{\AA} and 1400~\text{\AA} indices, associated with Si~\textsc{iii} and Si~\textsc{iv}, are also sensitive to the combined effects of stellar photospheric absorption, interstellar absorption, and nebular emission.

Building on the work of \citet{leitherer2011ultraviolet} and \citet{faisst2016rest}, \citet{vidal2017modelling} assessed the contamination affecting commonly used UV indices and identified the 1425~\text{\AA} and 1719~\text{\AA} indices as among the most robust tracers of stellar metallicity. The 1719~\text{\AA} index is particularly useful because it contains several medium- to high-ionisation transitions, including N~\textsc{iv} at 1718.6~\text{\AA}, Si~\textsc{iv} at 1722.5 and 1727.4~\text{\AA}, and Al~\textsc{ii} and Fe~\textsc{iv} transitions across the 1705--1729~\text{\AA} interval.

More recent stellar population synthesis models have also been used to calibrate UV indices as metallicity diagnostics for galaxies at $2 \lesssim z \lesssim 5$ \citep{Calabro21}. In particular, the 1501~\text{\AA} and 1719~\text{\AA} indices are largely insensitive to variations in the initial mass function, stellar age, dust attenuation, and nebular continuum emission when modelled with the \texttt{Starburst99} population synthesis code \citep{Leitherer1999}.

\begin{table}
    \centering
    \normalsize
    \begin{tabular}{c|c|c|c}
         \hline
         Index & \( \lambda_1 (\text{\AA})\) & \( \lambda_2 (\text{\AA})\) & Associated Abundance \\
         \hline
         1370 & 1360 & 1380 & Si \textsc{iii} \\
         1400 & 1385 & 1410 & Si \textsc{iv} \\
         1425 & 1413 & 1435 & Fe \textsc{v}, C \textsc{iii}, Si \textsc{iii} \\
         1460 & 1450 & 1470 & Ni \textsc{ii} \\
         1501 & 1496 & 1506 & S \textsc{v} \\
         1533 & 1530 & 1537 & Si \textsc{ii} \\
         1550 & 1530 & 1560 & C \textsc{iv} \\
         1719 & 1705 & 1729 & N \textsc{iv}, Si \textsc{iv}, Al \textsc{ii}, Fe \textsc{iv} \\
         1853 & 1838 & 1858 & Al \textsc{ii}, Al \textsc{iii}, Fe \textsc{ii} \\
         \hline
    \end{tabular}
    \caption{
    Summary of the UV indices examined in this study. The second and third columns list the wavelength boundaries $\lambda_1$ and $\lambda_2$, respectively, corresponding to the integration limits defined in Equation \ref{eqn:EW} for the measurement of equivalent widths. The fourth column indicates the associated abundance for each of the UV indices.
    }
    \label{tab:Indices Table}
\end{table}

\begin{table}
    \centering
    \begin{tabular}{c|c|c|c|c}
         \hline
         Index & Blue \(\lambda_1 (\text{\AA})\) & Blue \(\lambda_2 (\text{\AA})\) & Red \(\lambda_1 (\text{\AA})\) & Red \(\lambda_2 (\text{\AA})\) \\
         \hline
         1370 & 1345 & 1354 & 1436 & 1447 \\
         1400 & 1345 & 1354 & 1436 & 1447 \\
         1425 & 1345 & 1354 & 1436 & 1447 \\
         1460 & 1436 & 1447 & 1482 & 1491 \\
         1501 & 1482 & 1491 & 1583 & 1593 \\
         1533 & 1482 & 1491 & 1583 & 1593 \\
         1550 & 1482 & 1491 & 1583 & 1593 \\
         1719 & 1675 & 1684 & 1751 & 1761 \\
         1853 & 1797 & 1807 & 1871 & 1883 \\
         \hline
    \end{tabular}
    \caption{The blueward and redward pseudo-continuum windows used to estimate the local continuum level, selected to avoid contamination from nearby absorption features.}
    \label{tab:Table Pseudo}
\end{table}

\subsection{Metallicity} 
\begin{figure*}
    \centering
    \includegraphics[width=0.9\linewidth]{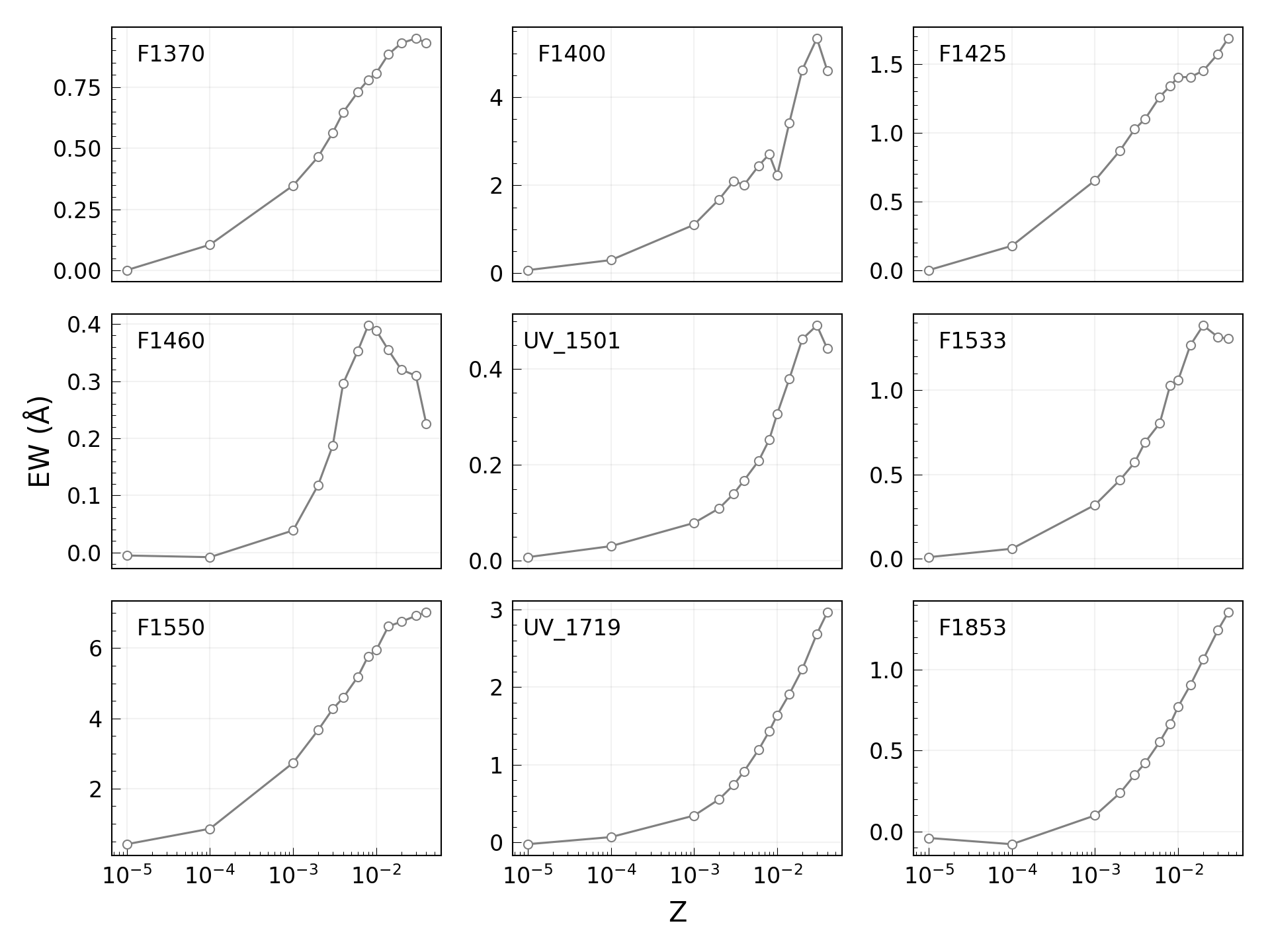}
    \caption{
    Equivalent width of UV indices as a function of stellar metallicity. The models span a metallicity range of $1 \times 10^{-5} \le Z \le 0.04$, assuming a continuous star formation history of 100 Myr.
    }
    \label{fig:Baseline}
\end{figure*}

The stellar population models used in this work are drawn from the BPASS v2.3 grid, which samples a discrete set of stellar metallicities spanning $10^{-5} \leq Z \leq 0.04$. The spectra are sampled at a $0.5~\text{\AA}$ fixed resolution defined as $\Delta\lambda =  $0.5~\text{\AA}, providing a consistent grid resolution across the full metallicity range. This range extends from extremely metal-poor populations to high metallicities and therefore covers the values expected for both chemically primitive and more evolved high-redshift galaxies. Since the grid is discrete in metallicity, equivalent-width trends are evaluated at the native BPASS metallicity points, with no extrapolation beyond the defined range.

The interpretation of the lowest- and highest-metallicity models requires some caution. At very low metallicity, stellar evolution tracks, atmosphere models, and line strengths are less well constrained observationally, while at the highest metallicities, line blanketing, blending, and saturation effects become increasingly important in the rest-frame UV. In addition, the BPASS grids assume fixed abundance patterns at each metallicity, with variations in individual elemental abundance ratios not being explicitly captured. The results should therefore be interpreted as predictions within the BPASS modelling framework, rather than as a complete description of all possible stellar abundance patterns. Nevertheless, the broad metallicity coverage of the grid provides a suitable basis for assessing the relative sensitivity of UV indices to stellar metallicity across the range relevant to the \flares\ galaxies considered in this work.

In this section, we examine the dependence of the UV indices listed in Table \ref{tab:Indices Table} and illustrated in Figure \ref{fig:Figure Pseudo} on stellar metallicity, Z, following the equivalent-width approach adopted in previous studies \citep{Rix04, Sommariva2012stellar, Calabro21}. The equivalent width (EW) of each index is calculated as

\begin{equation} \label{eqn:EW}
    EW_i = \int_{\lambda_{1}}^{\lambda_{2}} \left(1 - \frac{f(\lambda)}{f_{\mathrm{cont}}(\lambda)} \right) \, d\lambda
\end{equation}

With this convention, absorption features have positive equivalent widths, while emission features have negative values. Here, $f(\lambda)$ is the spectral flux density across the feature, $f_{\rm cont}(\lambda)$ is the local continuum flux density, and $\lambda_1$ and $\lambda_2$ define the wavelength bounds of the index.

Accurate EW measurements require an estimate of the local continuum that is minimally affected by line contamination. We therefore adopt the blueward and redward pseudo-continuum windows refined by \citet{Calabro21}, which are listed in Table \ref{tab:Table Pseudo}. These windows broaden the earlier definitions from $\approx 3-4~\text{\AA}$ to $\approx 10~\text{\AA}$, improving the continuum estimate around each UV index.

We generate synthetic spectra with \texttt{Synthesizer} using simple stellar populations from the BPASS v2.3 incident grid \citep{Byrne21}. Here, incident refers to the pure stellar spectrum before nebular reprocessing or radiative transfer effects are applied. This choice isolates the stellar absorption features that define the UV indices, allowing their metallicity dependence to be examined without additional contributions from nebular emission.

We use these grids to examine the relationship between equivalent width and stellar metallicity for each UV index, assuming a continuous star formation history of 100 Myr. As shown in Figure~\ref{fig:Baseline}, most indices exhibit a monotonic increase in equivalent width with metallicity. This trend is expected because higher metallicities increase line blanketing and opacity in stellar atmospheres, strengthening metal absorption features in the emergent UV spectrum.

The strong metallicity dependence of several indices indicates that the \texttt{Synthesizer} implementation captures the expected behaviour of the BPASS simple stellar population models. However, the response is not uniform across all indices. Figure~\ref{fig:Baseline} shows clear departures from monotonic behaviour for the 1400~\text{\AA} and, more prominently, the 1460~\text{\AA} indices. The 1400~\text{\AA} feature, commonly associated with Si~\textsc{iv}, exhibits a mild decrease in equivalent width near $Z \approx 10^{-2}$. This may reflect changes in the ionisation balance of silicon, line saturation, or the relative contribution of stellar photospheric and wind features as metallicity increases.

The 1460~\text{\AA} index shows a more pronounced deviation, with the equivalent width decreasing sharply at the highest metallicities. This feature is primarily associated with Ni~\textsc{ii} absorption and is therefore more directly tied to Fe-peak behaviour than the other UV indices considered here. The visible trend is not seen with the same strength in the remaining indices, likely due to various indices being dominated by multiple elements and some consisting of broader blends in which contributions from multiple transitions partly average over such behaviour.

The drop at high metallicity should therefore be interpreted with caution. The probable cause is a model-dependent feature of the BPASS v2.3 spectra used in this work, which are evaluated at a $0.5~\text{\AA}$ fixed resolution. This differs from the lower-resolution grids used by \citet{Byrne21}, where the same feature is not reported. The higher spectral sampling adopted here may make the 1460~\text{\AA} index more sensitive to narrow line structure, blending, and pseudo-continuum placement at the metal-rich end of the grid. As a result, the decline may reflect a combination of physical sensitivity to Fe-peak absorption and model-specific effects associated with the high-metallicity BPASS predictions. We therefore treat the 1460~(\text{\AA}) index as less reliable as a stand-alone metallicity diagnostic at high Z, particularly where emission modelling and spectral sampling can influence the measured equivalent width.

These departures show that, although many UV indices strengthen with increasing metallicity, individual features can exhibit non-linear responses to the detailed physical conditions of the stellar population. Such behaviour must be accounted for when applying UV indices to infer the chemical enrichment of high-redshift galaxies.

\subsection{Star Formation History}
\begin{figure*}
    \centering
    \includegraphics[width=0.9\linewidth]{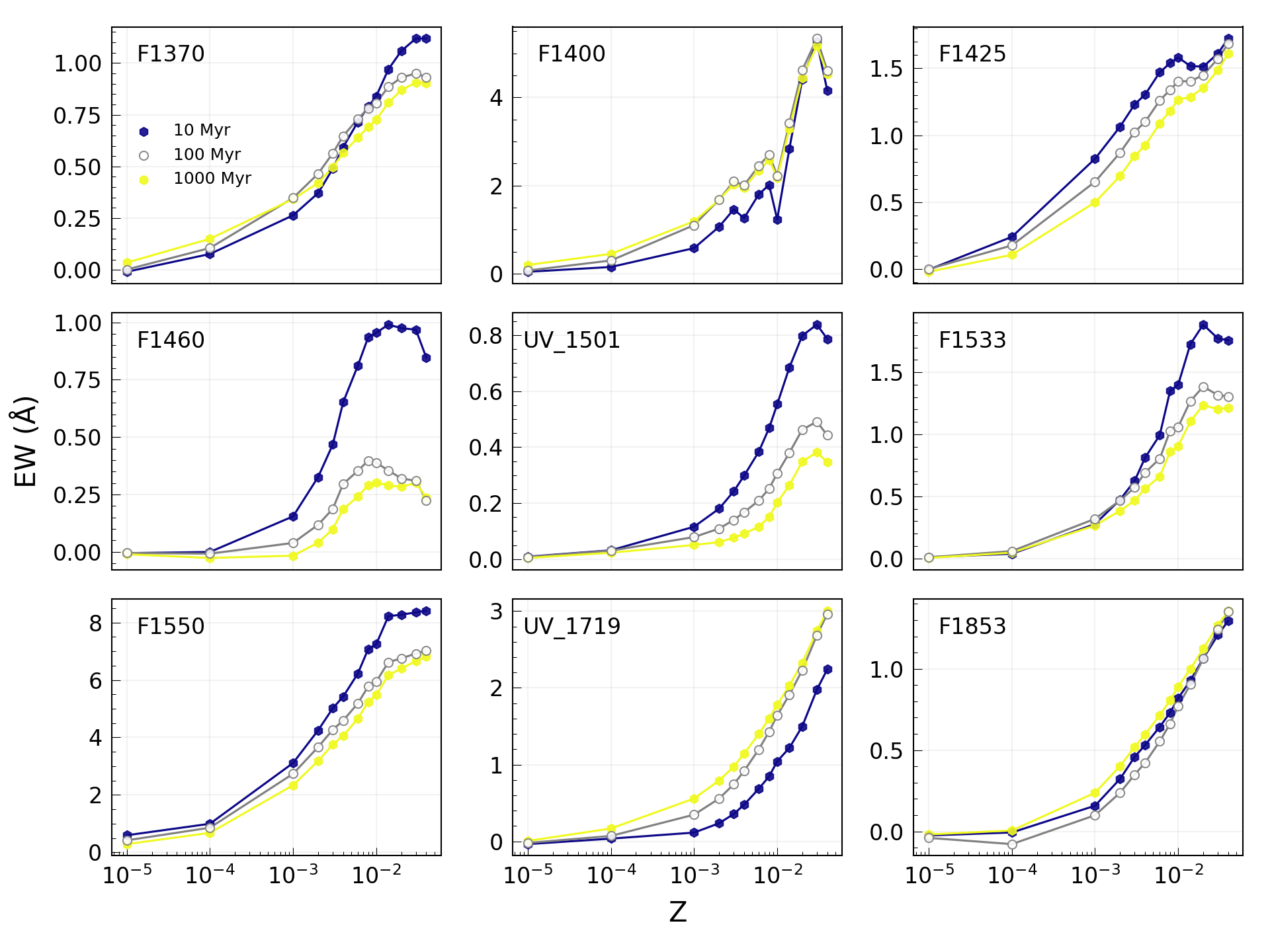}
    \caption{Equivalent width values of UV indices plotted as a function of wavelength for three distinct star formation histories. The spectra correspond to continuous star formation durations of 10 Myr (blue), 100 Myr (grey), and 1 Gyr (yellow).}
    \label{fig:SFH}
\end{figure*}

To assess the sensitivity of UV indices to star formation history, we construct three \texttt{Synthesizer} model spectra using BPASS single-metallicity stellar populations. Following \citet{Byrne21}, each model assumes continuous star formation over a different timescale, being 10~Myr, 100~Myr, or 1~Gyr. This allows the equivalent widths of the UV indices to be compared across populations with different characteristic stellar ages, providing a test of how recent star formation affects the absorption features.

Figure~\ref{fig:SFH} shows that the youngest populations generally produce stronger absorption across several indices than models with more extended star formation histories. However, the response varies between features, indicating that the sensitivity to star formation history depends on the ionisation state, metallicity, and age distribution of the contributing stellar populations.

The 1370~\text{\AA} index shows a metallicity-dependent response to star formation history. The 10~Myr model produces slightly weaker absorption than the 100~Myr model at $Z \lesssim 10^{-2}$, but stronger absorption at higher metallicity. The 1~Gyr model follows the opposite behaviour, producing comparatively stronger absorption at low metallicity and weaker absorption at high metallicity. The 1400~\text{\AA} index remains relatively stable across the three star formation histories, with only a small suppression in the 10~Myr model before converging with the 100~Myr case near $Z \approx 10^{-2}$. The 1425~\text{\AA} index shows similarly weak sensitivity, with only mild offsets between the three timescales.

Several indices respond more strongly to recent star formation. The 1460~\text{\AA} feature shows the largest enhancement, with the 10~Myr model producing equivalent widths up to a factor of $\approx$ 2.5 higher than the longer-duration cases. This likely reflects the greater contribution of young, massive stars to the UV spectrum. The 1501~\text{\AA} and 1550~\text{\AA} indices also strengthen in the 10~Myr model, consistent with increased sensitivity to young stellar populations. By comparison, the 1~Gyr spectra remain close to the 100~Myr baseline for these features. The 1533~\text{\AA} index is mostly unchanged across the three star formation histories at $Z \lesssim 10^{-3}$, becoming more age-dependent at higher metallicity.

The 1719~\text{\AA} blend behaves differently, showing a reduction in equivalent width for the 10~Myr model from $Z \approx 10^{-4}$ onwards, while the 1~Gyr model remains close to the 100~Myr case. The 1853~\text{\AA} feature is almost invariant, with only marginal differences between the three star formation histories. 

Overall, these results show that the response of UV indices to star formation history is highly index-dependent. While features such as 1460~\text{\AA}, 1501~\text{\AA}, and 1550~\text{\AA} strengthen in younger populations, the 1400~\text{\AA} and 1719~\text{\AA} indices weaken, indicating that different transitions respond differently to the changing age distribution and ionisation conditions of the stellar population.

\subsection{Initial Mass Function}
\begin{figure*}
    \centering
    \includegraphics[width=0.9\linewidth]{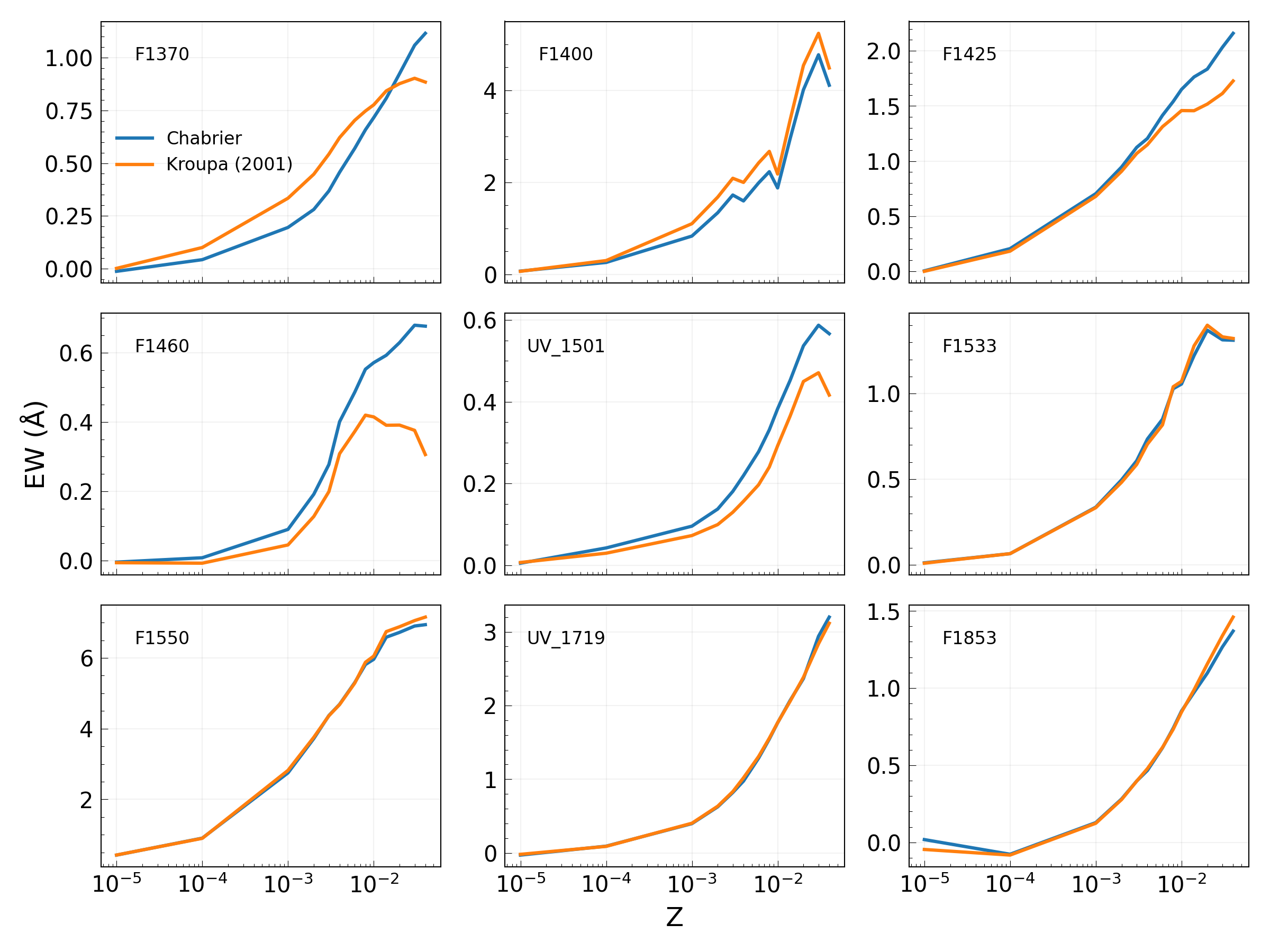}
    \caption{Equivalent widths of the UV indices computed using the two IMF choices available in the BPASS grids. The blue curve shows the Chabrier IMF, while the orange curve shows the Kroupa IMF adopted for the remainder of this work.}
    \label{fig:IMF}
\end{figure*}

The initial mass function (IMF) determines the relative contribution of stars of different masses to an integrated stellar population. Since rest-frame UV spectra are strongly influenced by young, massive stars, the adopted IMF can affect the predicted equivalent widths of UV absorption indices. In this section, we compare the two IMF choices available within the BPASS grids used in this work to assess whether the adopted IMF introduces significant differences in the UV indices.

We compare models generated using the Chabrier IMF \citep{Chabrier2003} and the \citet{kroupa2001variation} IMF. The Chabrier IMF adopts a log-normal form at low stellar masses and transitions to a power law at higher masses, while the \citet{kroupa2001variation} IMF adopts a piecewise power-law form. In the BPASS implementation used here, the \citet{kroupa2001variation} IMF has slopes of -1.30 over 0.1--$0.5~\mathrm{M}^*{\odot}$ and -2.35 at higher masses, extending to a maximum initial stellar mass of $300~\mathrm{M}^*{\odot}$. The purpose of this comparison is therefore to determine whether adopting the \citet{kroupa2001variation} IMF, used for the remainder of the analysis, has a substantial impact relative to the Chabrier alternative.

Figure~\ref{fig:IMF} shows that the largest differences between the Chabrier and \citet{kroupa2001variation} IMF occur for the 1370~\text{\AA} and 1460~\text{\AA} indices, associated with Si~\textsc{iii} and Ni~\textsc{ii}, respectively. These features are sensitive to the contribution of young, massive stars and to local line blending, making them more responsive to changes in the relative weighting of stellar masses. The 1460~\text{\AA} index remains more difficult to interpret because of its sensitivity to iron-peak absorption and pseudo-continuum placement.

More modest differences are present for the 1400~\text{\AA} and 1501~\text{\AA} indices. The 1400~\text{\AA} index generally exhibits higher equivalent widths for the \citet{kroupa2001variation} IMF, while the 1501~\text{\AA} index is stronger for the Chabrier IMF. These opposing trends indicate that the response to the IMF choice is index-dependent and is controlled by the specific transitions contributing to each feature, rather than by a uniform shift across the full UV index suite.

At wavelengths longer than 1550~\text{\AA}, the Chabrier and \citet{kroupa2001variation} IMF models produce nearly indistinguishable equivalent widths. This indicates that most of the longer wavelength UV indices are only weakly affected by the IMF choice considered here. Overall, the comparison shows that adopting the \citet{kroupa2001variation} IMF does not introduce major systematic differences for the majority of indices. We therefore use the \citet{kroupa2001variation} IMF for the remainder of the analysis, while noting that selected shorter-wavelength features, particularly 1370~\text{\AA}, 1460~\text{\AA}, 1400~\text{\AA}, and 1501~\text{\AA}, retain some sensitivity to the IMF choice.

\begin{figure*}
    \centering
    \includegraphics[width=0.9\linewidth]{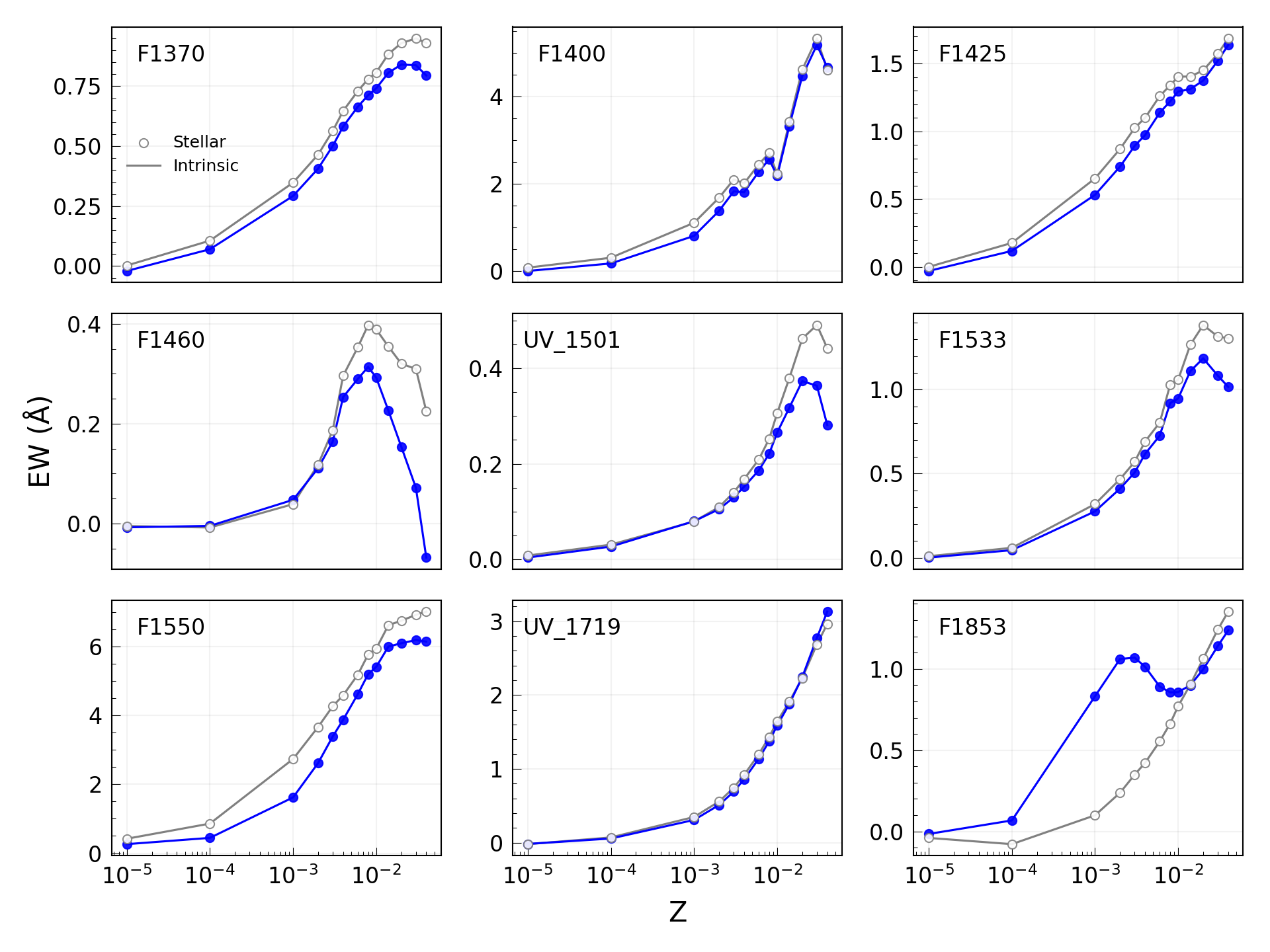}
    \caption{Equivalent width values of UV indices derived from both pure stellar spectra (grey) and intrinsic spectra (blue).}
    \label{fig:Intrinsic}
\end{figure*}

\subsection{Emission Modelling}\label{sec:intrinsic}
To assess the influence of nebular emission on UV index measurements, we compare pure stellar incident spectra with spectra that include nebular emission. In this section, incident refers to the stellar spectrum before nebular reprocessing, while intrinsic refers to the combined stellar and nebular spectrum before dust attenuation or IGM absorption are applied. The purpose of this comparison is therefore to isolate how the addition of nebular emission modifies the equivalent widths measured from otherwise identical stellar population models.

Nebular emission is incorporated into the BPASS model grids using the \texttt{CLOUDY} photoionisation code \citep{Cloudy23}. The BPASS continuous star formation models are coupled to a constant-density spherical gas geometry with $n_{\mathrm{H}} = 316~\mathrm{cm^{-3}}$, Orion-type dust grains, a reference ionisation parameter of $U = 0.01$, and Galactic Concordance abundance scaling. Depletion is included using the prescriptions of \citet{Jenkins2009} and \citet{Cloudy23}, scaled by a factor of 0.5. The models also include cosmic ray heating, a turbulence velocity of $100~\mathrm{km~s^{-1}}$, and a temperature floor of $100~\mathrm{K}$. Calculations are performed with \texttt{CLOUDY} v23.01 for a reference metallicity of Z = 0.01 at an age of $1~\mathrm{Myr}$, producing both continuum and standard emission-line outputs \citep{lovell2025synthesizer}. These grids allow the UV indices to be measured consistently with and without nebular emission.

Figure~\ref{fig:Intrinsic} shows that adding nebular emission modifies the measured equivalent widths, particularly at higher metallicity. For most UV indices, the differences between the pure stellar and stellar-plus-nebular spectra are small. The main exceptions are the 1460~\text{\AA} and 1853~\text{\AA} indices, which show stronger sensitivity to the adopted emission model and may be affected by changes to the local continuum or contamination in their pseudo-continuum regions.

The 1853~\text{\AA} index exhibits a notable increase in equivalent width when nebular emission is included. Although the same pseudo-continuum definitions as \citet{Calabro21} are adopted, this wavelength range is likely affected by line blanketing and blended absorption from multiple metal transitions. The enhanced equivalent width therefore suggests that the 1853~\text{\AA} feature is sensitive to the combined effects of metallicity, nebular reprocessing, and the adopted local continuum definition.

The 1460~\text{\AA} index shows the opposite response, with a pronounced decrease in equivalent width when nebular emission is included. This behaviour may arise from contamination of the pseudo-continuum windows or from changes to the local continuum introduced by nebular emission. Nebular continuum emission can partially fill absorption features, reducing the apparent absorption strength and, in some cases, producing a net emission-like feature. The 1460~\text{\AA} index is therefore highly sensitive to the adopted emission model and should be interpreted with caution.

Most other indices, including 1501~\text{\AA}, 1533~\text{\AA}, and 1550~\text{\AA}, show only small changes in equivalent width when nebular emission is included, with subtle differences becoming more apparent at higher metallicity. This suggests that nebular emission has a limited effect on these features, although changes to the local UV continuum can still influence their measured absorption strengths in metal-rich models.

These predictions represent the stellar and nebular contribution to the UV indices, however, they do not explicitly include absorption from the interstellar medium. This is an important limitation for comparison with observations, since several rest-frame UV indices can contain blended stellar and ISM absorption, particularly in high-redshift \textit{JWST}/NIRSpec spectra where individual components may not be fully separable. The model predictions should therefore be interpreted as a stellar-population baseline against which observed indices can be compared. In practice, applying these diagnostics to observations requires forward modelling that accounts for nebular emission, ISM absorption, spectral sampling, noise, and continuum placement. Indices with minimal ISM sensitivity are expected to provide the most direct stellar metallicity constraints, while indices affected by blended stellar and interstellar absorption should be used in combination with other features or treated with additional modelling.

These results indicate that the interpretation of UV indices requires careful consideration of various physical processes. While simplified stellar population models allow significant exploration into the underlying sensitivities of individual indices, they do not fully capture the complexity of real galaxies. Therefore, this analysis must be extended to the \flares\ hydrodynamical simulations, enabling the behaviour of UV indices to be examined within realistic galaxy environments that incorporate star formation and chemical enrichment histories.

\section{Predictions}\label{sec:predictions}
In this section, we present predictions for rest-frame UV indices using galaxies from the First Light and Reionization Epoch Simulations (\flares) at z $\gtrsim$ 5. In contrast to the simple stellar populations derived from \texttt{Synthesizer} models, \flares\ provides galaxies with self-consistent star formation and chemical enrichment histories. This allows the behaviour of UV indices to be examined across a realistic population of galaxies spanning a broad range of stellar masses, metallicities, and star formation properties.

Using these simulated galaxies, we investigate the relationship between UV index equivalent widths and the physical properties that regulate chemical enrichment in \flares. In particular, we examine the stellar mass--metallicity relation, the connection between metallicity and star formation, and the effect of adopting different stellar metallicity definitions, including mass-weighted, luminosity-weighted, and young-star-weighted metallicities.

\subsection{The First Light and Reionization Epoch Simulations}
In this work we utilise the First Light and Reionization Epoch Simulation \citep{FLARES-I,FLARES-II} suite to study the UV indices. In short, \flares\ is a suite of zoom re-simulations employing the AGNdT9 variant of the \eagle\ reference model \citep{schaye2015eagle, crain2015_eagle}, including adopting the same particle mass. The core \flares\ suite consists of 40 $14/h\ {\rm cMpc}$ radius re-simulations, using regions selected from a large $(3.2\ {\rm cGpc})^3$ low-resolution dark matter only simulation \citep{CEAGLE}. The chosen regions cover a wide range in over-density, $\delta + 1 \approx -1 \to 1$ (at $z \approx 4.7$), with an intentional over-representation of the highest-density environments. This selection strategy enables the simulation of galaxies spanning a broader range in stellar mass, luminosity, and environment than would be achievable using a periodic volume with comparable computational resources. To ensure that this environmental bias does not distort statistical inferences, each region is assigned a weight corresponding to its cosmological volume, allowing the combined sample to recover representative population-level properties. All subsequent analyses and distributions presented in this work are computed using these weights.

The resulting galaxy sample, covering stellar masses in the range $M_{\star} = 10^8$--$10^{11}\,{\rm M_{\odot}}$ (at $z = 5$), aligns well with all but the most sensitive \(\jwst\) observations \citep[]{FLARES-V}. Stellar properties are typically measured within a fixed spherical 30\,pkpc aperture, centred on the potential minimum of each galaxy, ensuring consistency in derived quantities such as stellar mass and metallicity while minimising contamination from surrounding material. For a more detailed introduction to \flares\ see \citet{FLARES-I}, and for the underlying physics model, \citet{schaye2015eagle} and \citet{crain2015_eagle}.

\subsubsection{Spectral Energy Distribution modelling}\label{sec:methods::sed}
We utilise synthetic spectral energy distributions (SEDs) generated for the \flares\ galaxies in this work. The original SED modelling produced for \flares\ is described in \citet{FLARES-II} and builds upon the efforts of \citet{Wilkins2016b} and \citet{Bluetides_dust}. Recently, we have updated this modelling to utilise the new \texttt{Synthesizer} synthetic observations pipeline. This is identical to our previous approach in \cite{FLARES-II}, now carried out with the publicly available \texttt{Synthesizer} \cite[]{lovell2025synthesizer, roper2025synthesizer} package. 

In brief, the generation of \flares\ galaxy spectra requires the construction of SEDs using the BPASS v2.3 binary stellar population synthesis model. The adopted grid corresponds to a binary BPASS model with a broken power-law IMF over $0.1$--$300\,{\rm M_{\odot}}$, and slopes $-1.3$ and $-2.35$. Each star particle within a galaxy is assigned an SED based on its mass, age, and metallicity, with the combined contribution from all star particles representing the integrated stellar emission of the system. In real galaxies, however, this stellar radiation is subsequently reprocessed through interactions with gas and dust in the ISM, modifying the emergent spectrum. To account for this, ionisation-bounded H\,\textsc{ii} regions are modelled using \texttt{CLOUDY}, which is coupled to each star particle to simulate nebular emission \citep[]{Cloudy23}.

Active galactic nucleus (AGN) emission is not included in this framework. This assumption is justified by the relatively low stellar masses and high redshifts of the galaxies considered, for which AGN activity is expected to be either absent or sub-dominant to stellar processes in the UV range. As such, the spectral features analysed in this work are assumed to be dominated by stellar and nebular contributions.

\begin{figure*}
    \centering
    \includegraphics[width=\linewidth]{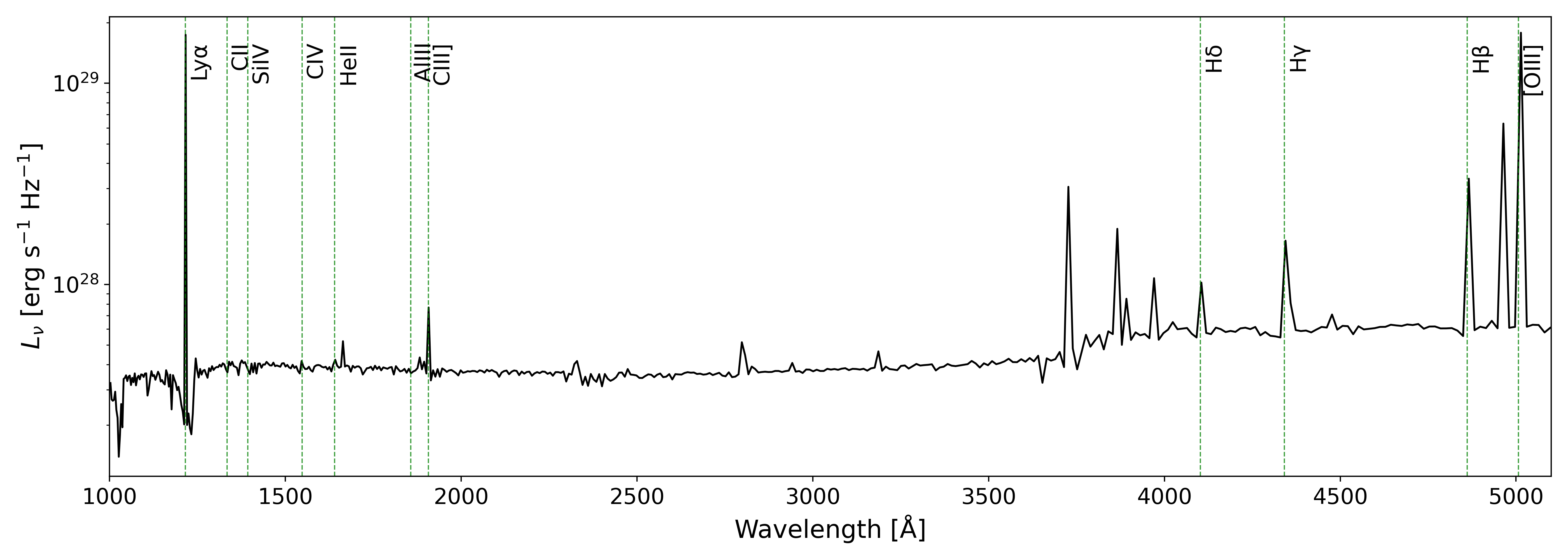}
    \caption{Sample rest-frame spectrum of a galaxy from the \flares\ simulation at $z=5$, showing prominent emission features such as Ly$\alpha$, C \textsc{iv}, He \textsc{ii}, H$\beta$, [O \textsc{iii}] that trace recent star formation and chemical enrichment in the early Universe.}
    \label{fig:FLARES-sample}
\end{figure*}

Assuming a spherical geometry and a nebular metallicity that matches the total metallicity of the corresponding star particle, the pure stellar spectrum is used as the incident radiation field. Additionally, a covering fraction of 1 is assumed, which corresponds to a LyC escape fraction of approximately zero for an ionization-bound nebula. The hydrogen density is set at $10^{2.5}$ cm$^{-3}$, and the ionization parameter is volume-averaged.

To model the impact of dust on the emergent spectra, we adopt a two-component attenuation prescription. First, a birth-cloud component is applied to all stars younger than 10 Myr, accounting for the dense, undispersed material that surrounds newly formed stellar populations. Second, attenuation by the diffuse interstellar medium is evaluated for every galaxy by computing the surface density of metals, $\Sigma_Z$, along the line of sight to each individual star particle.\footnote{By default the observer is placed along the $z$-axis, but see \citet{vijayan2024first} for an exploration of sight-line dependence.} The resulting metal column is converted to a wavelength-dependent optical depth, $\tau_\lambda$, using a dust--to--metal scaling fitted to the dust model implemented in \textsc{L-Galaxies} \cite[]{Vijayan2019}, which reproduces the lower dust-to-metal ratios inferred for high-redshift systems.

This approach yields a strongly environment-dependent UV luminosity function in which attenuation removes an average of $\sim40\%$ of the intrinsic UV light at $z = 7$ and grows increasingly dominant below $z \approx 6$.\footnote{In the far-UV, the attenuation obeys a nearly linear relation between observed and intrinsic luminosity; see \citet{FLARES-II}.}  Together, the birth-cloud and diffuse ISM components provide a self-consistent, particle-by-particle dust model that captures the heterogeneous obscuration expected in early galaxies while remaining consistent with empirical constraints on the dust content of the high-redshift Universe.

This process culminates in generating synthetic spectra for each galaxy, where the continuum of one such galaxy from the \flares\ simulation is illustrated in Figure \ref{fig:FLARES-sample}. The flux as a function of wavelength encodes the absorption and emission features arising from the underlying stellar populations and their interaction with the interstellar medium. In the simulations, each galaxy is constructed from a physically motivated set of stellar and gas properties assuming a continuous star formation history of 100 Myr, ensuring that the resulting spectra reflect realistic galactic conditions and evolutionary states.

\begin{figure}
    \centering
    \includegraphics[width=\columnwidth]{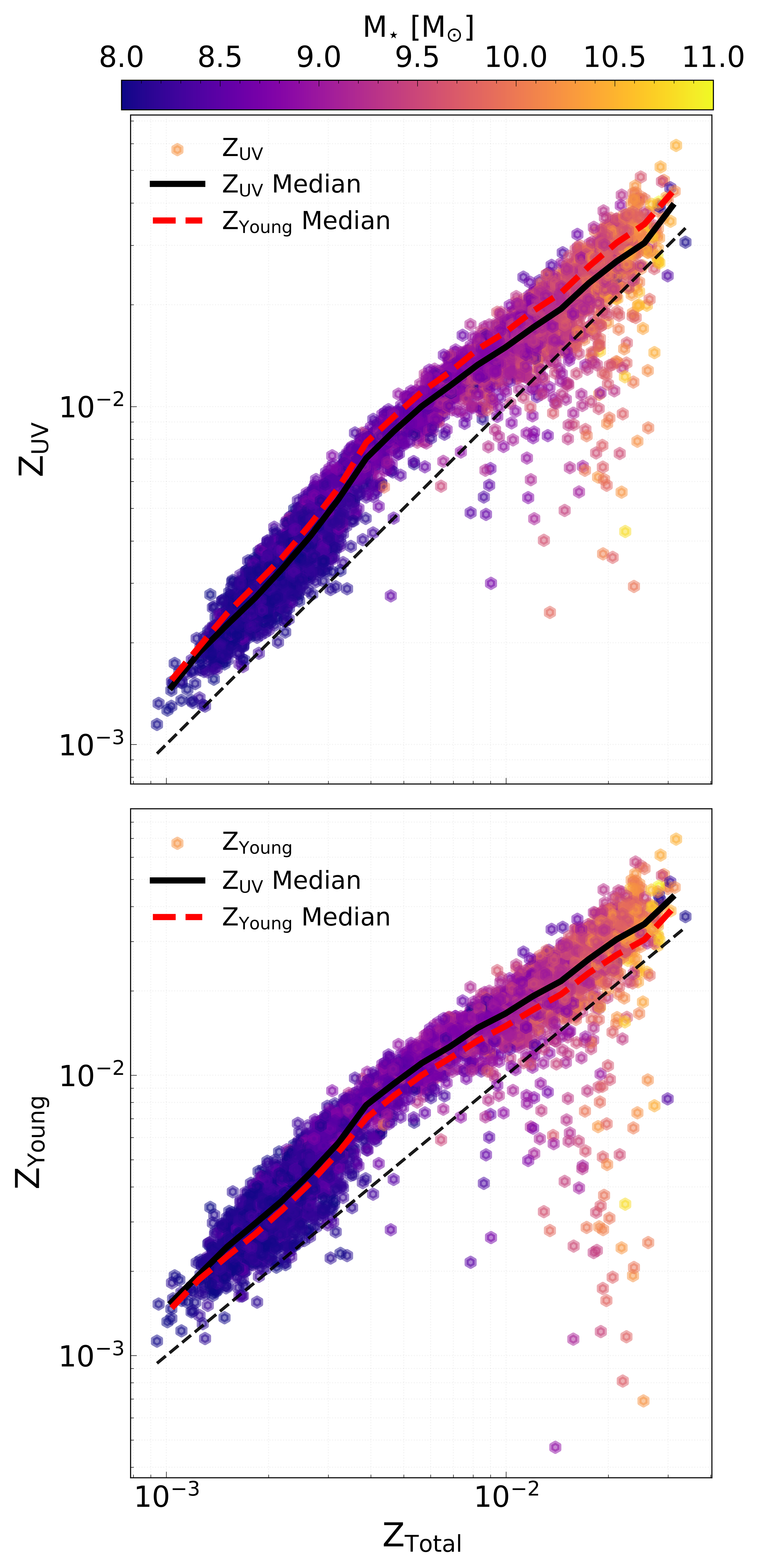}
    \caption{Comparison of metallicity measurements in \flares\ galaxies at z = 5. The top panel presents the UV luminosity-weighted metallicity plotted against the total mass-weighted stellar metallicity, while the bottom panel displays the mass-weighted metallicity of stars younger than 10 Myr versus the total mass-weighted stellar metallicity. The colour bar indicates the stellar mass of each galaxy, spanning from $10^8\,M_\odot$ (purple) to $10^{11}\,M_\odot$ (yellow).}
    \label{fig:Z_correlation}
\end{figure}

\subsection{Physical properties in \flares}
Interpreting metallicity in high-redshift galaxy simulations requires a consistent connection between stellar population properties and observable spectral features. \flares\ provides this connection by combining chemical enrichment, stellar evolution, and feedback with synthetic spectral modelling \citep{Wilkins2016a, FLARES-II}. This enables metallicity estimates to be compared directly with the UV spectral diagnostics.

\subsubsection{Metallicity Definitions}
In this section, we examine different definitions of stellar metallicity in \flares\ by comparing alternative weighting schemes with the total mass-weighted stellar metallicity. UV index equivalent widths are most sensitive to the stellar populations that dominate the emergent rest-frame UV light, rather than necessarily to the full stellar mass distribution. The adopted metallicity definition therefore affects how model predictions are interpreted. We assess whether total mass-weighted stellar metallicity provides an appropriate proxy for the populations shaping the UV absorption features, or whether luminosity or age-weighted definitions better trace the metallicity relevant to equivalent-width measurements.

Figure~\ref{fig:Z_correlation} presents these comparisons as a function of the total mass-weighted stellar metallicity. The total mass-weighted value, associated with both panels, is defined as

\begin{equation}
Z_{\mathrm{Total}} = \frac{\sum_{i} m_i Z_i}{\sum_{i} m_i},
\end{equation}

\noindent where $m_i$ and $Z_i$ denote the mass and metallicity of the $i$-th stellar particle, respectively. This quantity represents the metallicity averaged over the entire stellar population of a galaxy. While this global metric is widely used in simulation analyses \cite[][]{vogelsberger_introducing_2014, schaye2015eagle}, it does not necessarily trace the metallicity of the young or UV-bright stars that dominate the formation of UV absorption features.

The upper panel of Figure~\ref{fig:Z_correlation} shows the UV-luminosity-weighted metallicity, $Z_{\mathrm{UV}}$, as a function of the total mass-weighted stellar metallicity, $Z_{\mathrm{Total}}$. The UV-weighted metallicity is defined as

\begin{equation}
Z_{\mathrm{UV}} = \frac{\sum_{i} L_{i,\mathrm{UV}} Z_i}{\sum_{i} L_{i,\mathrm{UV}}},
\end{equation}

\noindent where $L_{i,\mathrm{UV}}$ and $Z_i$ denote the intrinsic UV luminosity and metallicity of the $i$-th stellar particle, respectively. This weighting directly traces the stellar populations responsible for shaping UV absorption indices. Each point corresponds to a single \flares\ galaxy, weighted by total stellar mass ($\sim10^8$--$10^{11}\,M_\odot$). A strong positive correlation is observed, with the binned median closely following the 1:1 relation. This behaviour indicates that, for the majority of \flares\ galaxies, the metallicity of UV-dominant stars broadly tracks the global stellar metallicity. Consequently, the use of $Z_{\mathrm{Total}}$ as an input when modelling UV equivalent widths provides a reasonable approximation in most cases.

The lower panel shows the mass-weighted metallicity of stars younger than 10~Myr, $Z_{\mathrm{young}}$, as a function of $Z_{\mathrm{Total}}$. This quantity is defined as

\begin{equation}
Z_{\mathrm{young}} = \frac{\sum_{i,\, t_i < 10\,\mathrm{Myr}} m_i Z_i}{\sum_{i,\, t_i < 10\,\mathrm{Myr}} m_i},
\end{equation}

\noindent where $m_i$ and $t_i$ represent the mass and age of each stellar particle. This metric isolates the chemical properties of the most recently formed stellar populations, which are expected to contribute significantly to UV line formation in actively star-forming systems. Although a positive correlation is again evident, the relation exhibits increased scatter and a systematic offset from the 1:1 line. This broader dispersion likely reflects short-timescale variations in star formation and localised enrichment processes within \flares\ galaxies. Such variations imply that equivalent widths tied most strongly to the youngest stellar populations may not be fully captured by $Z_{\mathrm{Total}}$ alone, particularly in systems experiencing rapid or bursty star formation.

Figure~\ref{fig:Z_correlation} shows that the alternative stellar metallicity definitions correlate with the total mass-weighted stellar metallicity, but with different levels of scatter and systematic offset. These differences depend on whether the weighting emphasises the full stellar population, the UV-bright component, or the youngest stars.

This has important consequences for interpreting UV equivalent widths in \flares. When the UV-luminosity-weighted metallicity closely follows the total mass-weighted value, the latter provides a reasonable proxy for the stellar populations dominating the rest-frame UV continuum. However, the larger scatter associated with the young stellar component indicates that UV indices sensitive to recent star formation may trace metallicities that differ from the global stellar average.

\subsubsection{The Mass-Metallicity Relation (MZR)}\label{sec:MZR}
The mass--metallicity relation (MZR) predicted by \flares\ provides a means of connecting stellar mass growth to chemical enrichment at high redshift. This analysis builds on the framework established for \textsc{Eagle} \citep{schaye2015eagle} and subsequent extensions made to early galaxy populations \citep{de2017galaxy,FLARES-VII,kotiwale2026rapid}. A central consideration in these studies is the distinction between different stellar metallicity definitions, including total mass-weighted, young-star mass-weighted, and luminosity-weighted metallicities. Each definition emphasises a different component of the stellar population and therefore traces a different aspect of the enrichment history. Comparing these quantities allows the relationship between chemical enrichment and observable galaxy properties to be assessed at $z \gtrsim 5$.

\begin{figure}
    \centering
    \includegraphics[width=\columnwidth]{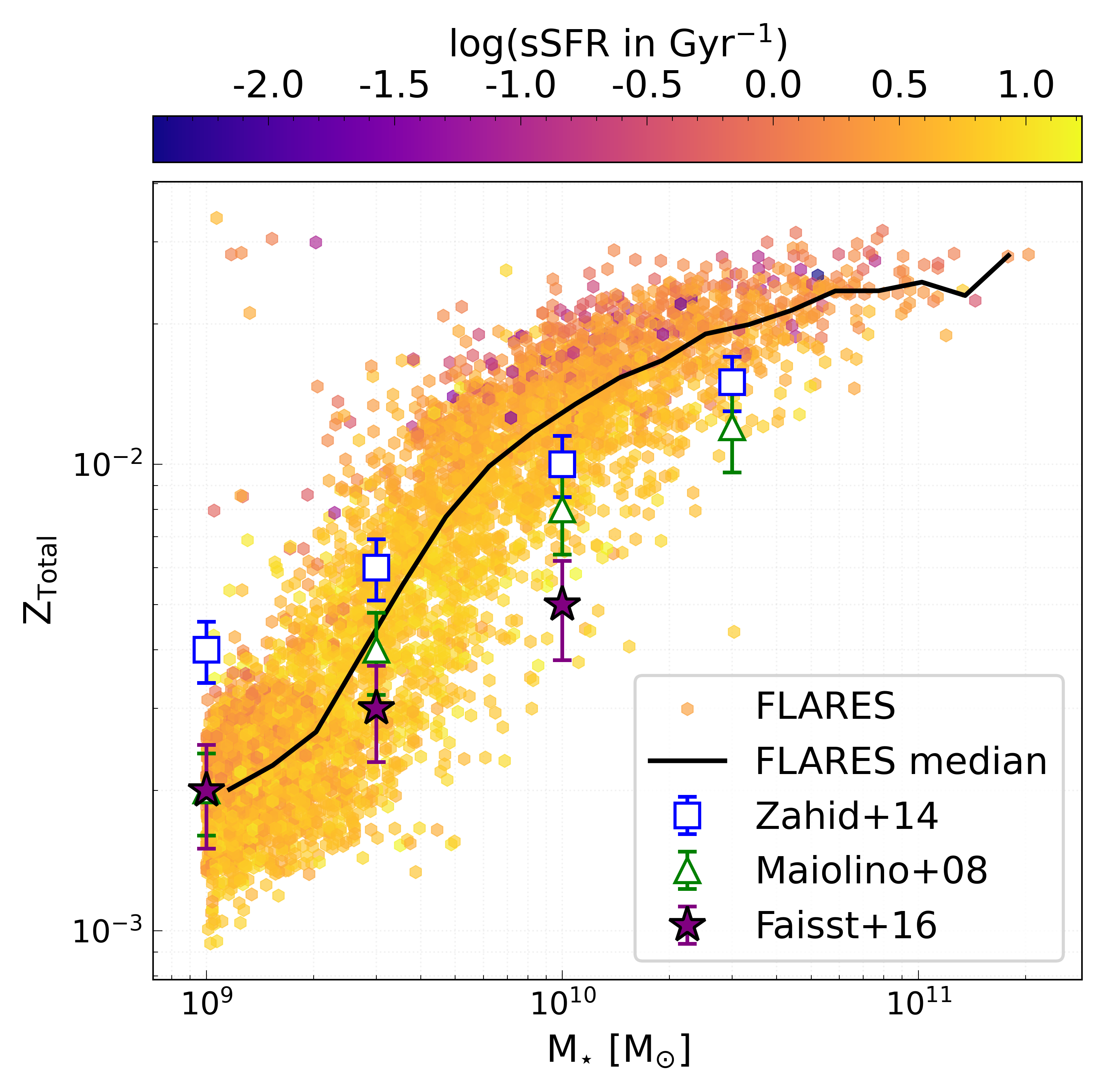}
    \caption{The total mass-weighted stellar metallicity as a function of stellar mass at $z = 5$ and $M_{\odot}\geq 10^{9}$, with individual data points weighted by specific star formation rate and a median line (black).}
    \label{fig:TotalMZR}
\end{figure}

Figure~\ref{fig:TotalMZR} shows the MZR derived using the total mass-weighted stellar metallicity as a function of stellar mass. Points are weighted by the specific Star Formation Rate (sSFR), calculated using the SFR averaged over 100~Myr, and the relation is compared with observational constraints at $z \approx 5$ \citep{Maiolino2008,Zahid2014,Faisst2016}. The \flares\ galaxies show a clear positive correlation between stellar mass and metallicity, with more massive systems reaching higher total stellar metallicities. This reflects the more efficient retention and recycling of enriched material in deeper gravitational potentials. The sSFR colour gradient indicates a secondary dependence on star formation activity, implying that low-mass, high-sSFR galaxies preferentially occupy the low-metallicity envelope, while more massive, lower-sSFR systems populate the metal-rich region. The running median follows the expected MZR form, with metallicities shifted to the lower values characteristic of galaxies at $z \gtrsim 5$.

\begin{figure}
    \centering
    \includegraphics[width=\columnwidth]{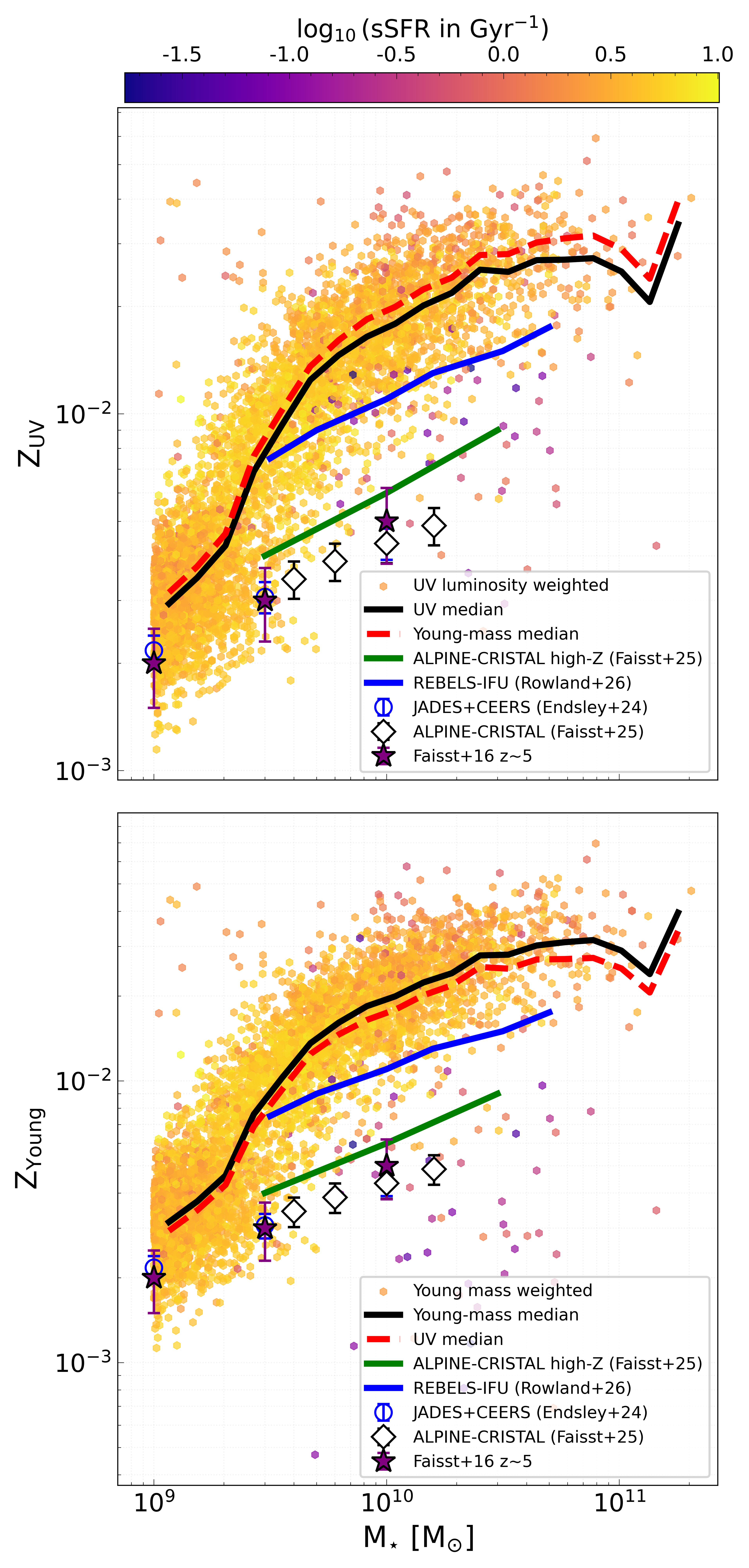}
    \caption{The mass metallicity relation for two distinct metallicity indicators, weighted specific star formation rate ($log(sSFR\ in\ Gyr^{-1})$). The top panel displays UV luminosity-weighted stellar metallicity, while the bottom panel shows the mass-weighted metallicity of stars younger than 10 Myr. The median relation is denoted by the black line.}
    \label{fig:UVMZR}
\end{figure}

Figure~\ref{fig:UVMZR} complements this global relation by comparing metallicities that emphasise different stellar components. The upper panel shows the UV luminosity-weighted stellar metallicity, which traces the populations dominating the emergent rest-frame UV light, while the lower panel shows the metallicity of stars younger than 10~Myr, tracing the most recent star formation. Both quantities are plotted against $\log_{10}(M_{*}/M_{\odot})$ and use the same sSFR weighting as Figure~\ref{fig:TotalMZR}. Although these definitions probe different stellar populations, both show increasing metallicity with stellar mass and lower metallicity at higher sSFR for fixed mass, consistent with recent observational and theoretical studies of high-redshift galaxies \citep{Faisst2016,endsley2024star,ALPINE_CRISTAL_JWST,ren2025alpine,Rowland2026_REBELS_IFU}. The young-star metallicity exhibits greater scatter, reflecting the stochastic nature of recent star formation, enrichment, and feedback. By contrast, the UV luminosity-weighted metallicity produces a tighter relation, indicating that the integrated UV light averages over short-timescale fluctuations and provides a more stable tracer of the metallicity relevant to UV spectral diagnostics.

Taken together, Figures~\ref{fig:TotalMZR} and~\ref{fig:UVMZR} show that different metallicity definitions trace complementary aspects of chemical enrichment in \flares. The total mass-weighted metallicity reflects the integrated enrichment history of the stellar population, the young-star metallicity traces the most recent enrichment events, and the UV luminosity-weighted metallicity emphasises the stars that dominate the emergent rest-frame UV spectrum. Having established these metallicity trends, we now examine individual UV indices in \flares\ to determine how their equivalent widths relate to the chemical properties of high-redshift galaxies.

\subsection{The UV Indices in \flares\ Simulations}
\begin{figure*}
    \centering
    \includegraphics[width=\textwidth]{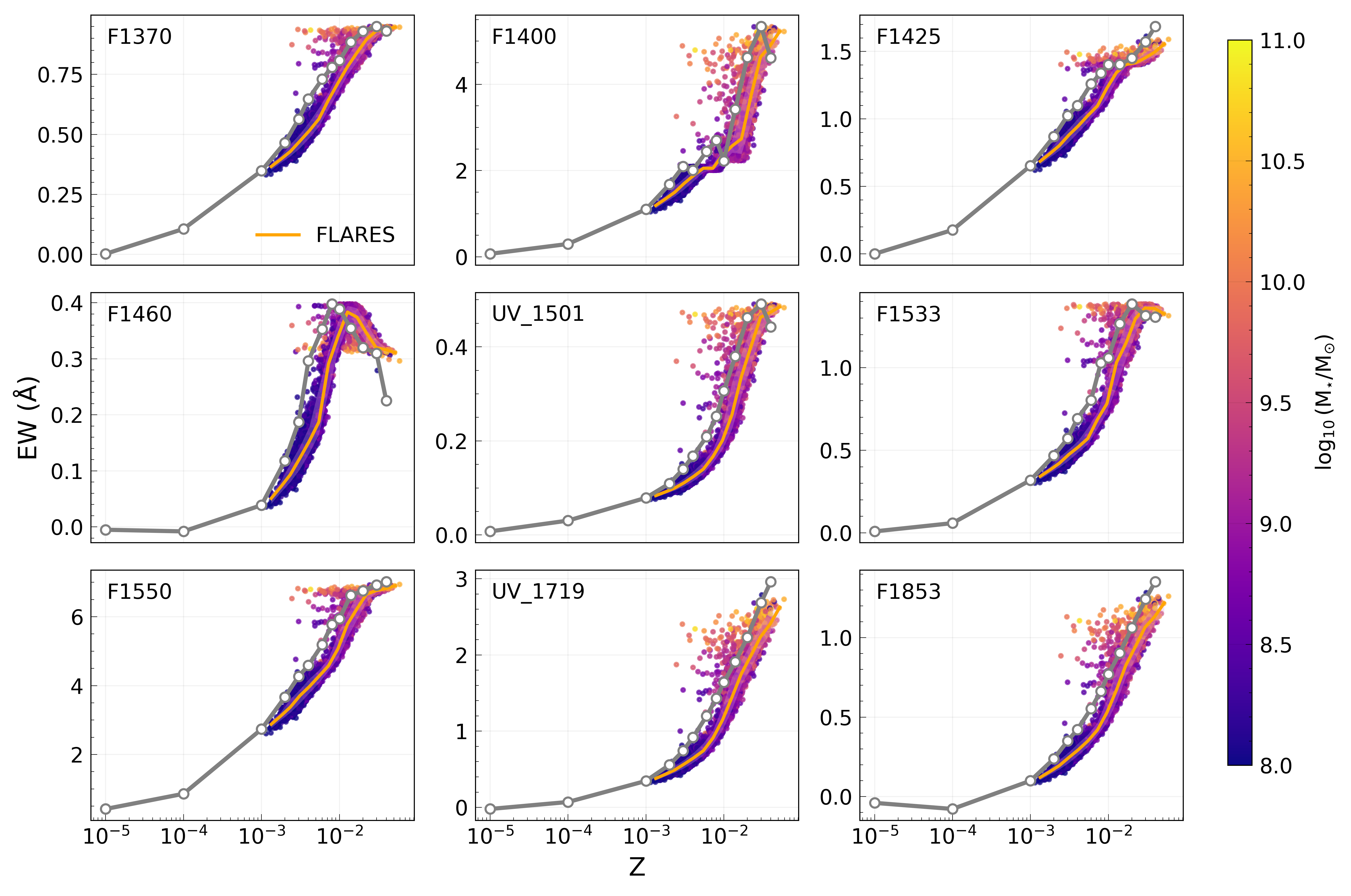}
    \caption{Comparison of UV absorption line equivalent widths between predictions from simple stellar populations and individual \flares\ galaxies at $z = 5$. The median relation highlights the correspondence between simulated galaxy spectra and model predictions as a function of UV luminosity–weighted stellar metallicity.}
    \label{fig:Flares_EW}
\end{figure*}

In \flares, galaxies are characterised by distributions of stellar metallicity rather than by a single fixed value. These distributions arise from their resolved star formation and chemical enrichment histories. As a result, equivalent widths are not evaluated at a prescribed metallicity, as in simple stellar population models, but are instead measured from galaxy-integrated spectra constructed from all contributing stellar particles. Each particle contributes according to its age, mass, and metallicity, so the composite spectrum encodes the time-dependent enrichment history of the galaxy.

Here, we restrict the analysis to the pure stellar component of the spectrum, resulting in the measured equivalent widths being driven by stellar absorption features. Although the full \flares\ suite can include nebular emission and dust attenuation, these effects are excluded in this part of the analysis to isolate the intrinsic stellar contribution to the UV indices.

This approach differs from the simple stellar population models presented in Section~\ref{sec:theory}. In those models, equivalent widths are measured from spectra defined by a single metallicity and a prescribed star formation history, allowing the response of each UV index to be examined under controlled conditions. In \flares, BPASS spectra are assigned to individual stellar particles according to their ages and metallicities, then summed to produce a galaxy-integrated spectrum before the equivalent widths are measured. The resulting UV absorption features therefore reflect both the current stellar metallicity distribution and the cumulative star formation and enrichment history of each galaxy.

Figure~\ref{fig:Flares_EW} shows that most UV indices increase in equivalent width with increasing stellar metallicity, broadly consistent with the trends found for the simple stellar population models. The slopes and scatter vary between indices, highlighting differences in their sensitivity to stellar age distributions, metallicity distributions, and the composite nature of the \flares\ spectra. Weighting galaxies by stellar mass reinforces the mass--metallicity relation discussed in Section~\ref{sec:MZR}, with more massive systems preferentially occupying the higher-metallicity, stronger-absorption regime.

The 1370~\text{\AA} and 1400~\text{\AA} indices show close agreement between the \flares\ predictions and the simple stellar population models. Some additional scatter is present at the high-metallicity end, particularly among more massive galaxies, with the overall equivalent-width--metallicity trend remaining well preserved. This suggests that these Si-dominated features remain sensitive to bulk stellar metallicity in composite galaxy spectra, although secondary effects from stellar age distributions and population mixing contribute to the observed dispersion.

The blended 1425~\text{\AA} feature also follows a broadly consistent trend with metallicity, but with modest scatter and some flattening at higher metallicities. This may indicate that the dominant absorption features begin to saturate, leading TO further increases in metallicity producing smaller changes in equivalent width. The composite nature of the index, which includes Fe-, C-, and Si-bearing transitions, may also contribute to the increased scatter by introducing sensitivity to stellar atmospheric conditions and age distributions.

The 1460~\text{\AA} index shows a clearer departure from monotonic behaviour, with flattening and, in some cases, a decline in equivalent width at $Z \gtrsim 10^{-2}$. This trend is most apparent among the most metal-rich and massive \flares\ galaxies at $z = 5$. As discussed above, the 1460~\text{\AA} feature is sensitive to the detailed behaviour of Fe-peak absorption and to the definition of the local pseudo-continuum. Its response in \flares\ therefore likely reflects a combination of metallicity, stellar population mixing, and changes in the local continuum shape, rather than a simple one-to-one dependence on stellar metallicity.

The 1501~\text{\AA} index broadly follows the simple stellar population predictions, although increased scatter and a small number of outliers appear at the high-mass end. A mild flattening is also present at higher metallicities, particularly in more massive systems. This behaviour suggests that the S~\textsc{v} dominated absorption remains metallicity sensitive, with the overall response becoming less direct in composite stellar populations where variations in stellar age and recent star formation contribute to the measured equivalent width.

The 1533~\text{\AA} and 1550~\text{\AA} indices recover the expected positive scaling with metallicity, with moderate dispersion at higher metallicities. The 1550~\text{\AA} index, associated with C~\textsc{iv}, shows mild flattening, consistent with its additional sensitivity to stellar winds and circumstellar absorption. These effects can reduce the extent to which the line traces metallicity alone, particularly in more complex stellar populations.

At longer wavelengths, the 1719~\text{\AA} and 1853~\text{\AA} indices remain broadly consistent with the simple stellar population predictions. Both indices show increased scatter toward higher metallicities, with more massive galaxies spanning a wider range of equivalent widths. This dispersion likely reflects the blended nature of these features, where multiple transitions contribute to the measured absorption and introduce sensitivity to the stellar age distribution and composite spectral structure of \flares\ galaxies.

Overall, the close agreement between \flares\ and simple stellar population based predictions supports the validity of UV absorption indices as tracers of stellar metallicity within fully cosmological simulations. While all indices retain sensitivity to metallicity, robustness varies. The 1425~\text{\AA} and 1550~\text{\AA} features emerge as the most stable diagnostics, whereas all other indices provide additional sensitivity to star formation variability and enrichment complexity within high-redshift galaxies.

\subsubsection{The redshift evolution of UV indices in \flares.}
The epoch of reionization spans a broad redshift interval, approximately $z \sim 10$ to $z \sim 6$, during which galaxies undergo rapid star formation and chemical enrichment. The \flares\ simulations allow the evolution of rest-frame UV absorption features to be examined across this period, linking changes in equivalent width to the time-dependent build-up of stellar populations and metals.

We measure UV index equivalent widths across discrete redshift snapshots spanning $5 \lesssim z \lesssim 10$ to assess whether the metallicity sensitivity of these diagnostics remains stable as galaxy populations evolve. For each redshift slice, we calculate the median equivalent width using binned, weighted quantile estimates. This approach captures the underlying galaxy population while reducing sensitivity to outliers and sampling fluctuations.

\begin{figure*}
    \centering
    \includegraphics[width=\textwidth]{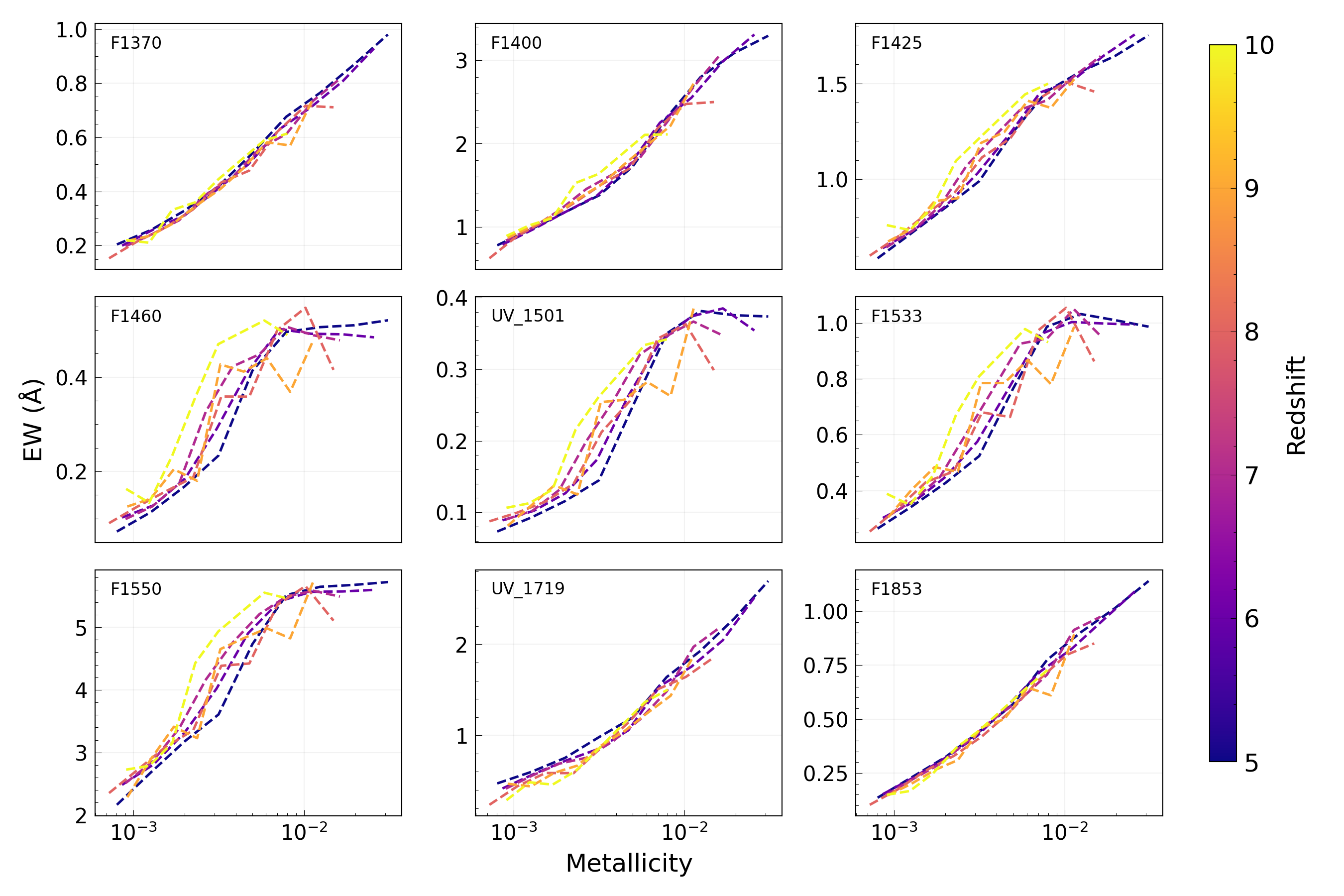}
    \caption{Evolution of UV absorption line equivalent widths as a function of stellar metallicity for galaxies in \flares\ over the redshift range $5 \le z \le 10$.}
    \label{fig:redshift-flares}
\end{figure*}

Across the UV index suite, the 1370~\text{\AA} and 1400~\text{\AA} features remain relatively stable with redshift. This behaviour is consistent with the trends discussed in Section~\ref{sec:theory}, where these Si-dominated absorption features show comparatively weak sensitivity to variations in star formation history. Although modest changes occur for the shortest star formation timescales, the overall response remains weaker than for several other indices. The limited redshift evolution in \flares\ therefore suggests that these features remain useful tracers of bulk stellar metallicity in composite high-redshift galaxy spectra.

The 1425~\text{\AA} index shows a modest increase in scatter across redshift, reflecting its blended nature and the contribution of Fe~\textsc{v}, C~\textsc{iii}, and Si~\textsc{iii} absorption. Since these transitions respond differently to stellar age, metallicity, and atmospheric conditions, the integrated equivalent width is more sensitive to the detailed stellar population than the simpler Si-dominated features.

The 1460~\text{\AA} and 1501~\text{\AA} indices show the strongest redshift-dependent variation in the sample. The equivalent widths of these indices exhibit increased scatter between snapshots, indicating sensitivity to the evolving stellar populations in \flares. The 1460~\text{\AA} feature, associated with Ni~\textsc{ii}, is already identified as non-monotonic in the simple stellar population models, while the 1501~\text{\AA} feature also shows sensitivity to recent star formation. These indices are therefore less robust as stand-alone metallicity tracers unless the star formation history and local continuum placement are carefully modelled.

The 1533~\text{\AA} and 1550~\text{\AA} indices show more moderate redshift variation. The 1550~\text{\AA} feature, associated with C~\textsc{iv}, remains metallicity-sensitive but is also affected by stellar winds and circumstellar absorption. These additional dependencies can broaden the equivalent-width distribution, particularly at the highest redshifts where galaxy populations are more bursty and less chemically mature.

The 1719~\text{\AA} and 1853~\text{\AA} indices are among the most stable features across the redshift range considered. The 1719~\text{\AA} blend, which includes contributions from N~\textsc{iv}, Si~\textsc{iv}, Al~\textsc{ii}, and Fe~\textsc{iv}, follows the overall metallicity trend with comparatively little redshift-dependent scatter. Its blended structure may average over variations in individual transitions, making the measured equivalent width less sensitive to short-timescale fluctuations in the stellar population. The 1853~\text{\AA} index also shows relatively stable behaviour, although its interpretation remains subject to the potential line-blending and continuum-placement effects discussed above.

A clear deviation appears at z = 9, particularly near $Z \sim 10^{-2}$. This is likely driven by small-number statistics rather than a population-wide change in the metallicity--index relation. In \flares, metal-rich galaxies are rare at $z \sim 9$, meaning that the median trends in this regime are dominated by a small number of massive, rapidly enriched systems. These galaxies are also more likely to experience bursty star formation, which can modify the UV continuum and increase the scatter in measured equivalent widths. The combination of sparse sampling and burst-driven variability therefore provides a plausible explanation for the increased scatter and occasional dips in equivalent width at this redshift.

Despite this deviation, the 1719~\text{\AA} index continues to follow the overall metallicity trend at z = 9. This supports the use of the 1719~\text{\AA} index as one of the more stable UV metallicity diagnostics in the \flares\ predictions. By contrast, narrower or more element-specific indices, such as 1460~\text{\AA} and 1501~\text{\AA}, show larger deviations, reinforcing the importance of index composition when assessing the reliability of UV absorption features across redshift.

Overall, these results show that the redshift dependence of UV indices in \flares\ is index-dependent. Features such as 1370~\text{\AA}, 1425~\text{\AA}, 1719~\text{\AA}, and, to a lesser extent, 1853~\text{\AA}, preserve relatively stable metallicity trends across $5 \lesssim z \lesssim 10$. Other indices, particularly 1460~\text{\AA} and 1501~\text{\AA}, are more sensitive to population level effects. These variations highlight the need to account for redshift-dependence when applying UV indices to galaxies during the epoch of reionization.

\section{Conclusions}\label{sec:conc}
In this work, we present predictions for the behaviour of rest-frame UV absorption indices using the \texttt{Synthesizer} synthetic spectra pipeline and the \flares\ hydrodynamical simulations. We use controlled stellar population models to examine the sensitivity of each index to metallicity, star formation history, and nebular emission, before extending the analysis to composite galaxy spectra from \flares. Our main conclusions are summarised as follows.

\begin{table*}
\centering
\begin{tabular}{lcccc}
\hline
\textbf{Index}
& \textbf{Metallicity}
& \textbf{SFH Variance}
& \textbf{Emission}
& \textbf{Redshift} \\
\hline

1370~\text{\AA}
& Monotonic
& Insensitive
& Minor
& Stable \\

1400~\text{\AA}
& Monotonic
& Insensitive
& Minor
& Stable \\

1425~\text{\AA}
& High-$Z$ plateau
& Insensitive
& Minor
& Weak variance \\

1460~\text{\AA}
& Non-monotonic, scattered
& Strong response
& Significant
& High variance \\

1501~\text{\AA}
& Monotonic
& Strong response
& Moderate at high-$Z$
& High variance \\

1533~\text{\AA}
& High-$Z$ plateau
& Moderate response
& Minor
& Moderate variance \\

1550~\text{\AA}
& Flattening at high $Z$
& Moderate response
& Moderate
& Moderate variance \\

1719~\text{\AA}
& Tight monotonic
& Moderate response
& Negligible
& Stable \\

1853~\text{\AA}
& Tight monotonic
& Insensitive
& Significant
& Stable \\
\hline

\end{tabular}
\caption{Summary of the sensitivity of each UV index to metallicity, star formation history, nebular emission, and redshift evolution in \texttt{Synthesizer} and \flares.}
\label{tab:UV_indices_conclusion}
\end{table*}

Table~\ref{tab:UV_indices_conclusion} summarises the index-specific behaviour identified in this work. Most UV indices show an increase in equivalent width with stellar metallicity, although the strength and linearity of this relation varies between features. The 1370~\text{\AA}, 1400~\text{\AA}, 1501~\text{\AA}, 1719~\text{\AA}, and 1853~\text{\AA} indices exhibit broadly monotonic behaviour, while the 1425~\text{\AA}, 1533~\text{\AA}, and 1550~\text{\AA} features show mild flattening or plateauing at higher metallicity. The 1460~\text{\AA} index displays the strongest departure from simple monotonic behaviour, with increased scatter and sensitivity to secondary effects.

The indices respond distinctly to star formation history. The 1460~\text{\AA}, 1501~\text{\AA}, 1533~\text{\AA}, and 1550~\text{\AA} indices are most sensitive to short-duration star formation, while the 1370~\text{\AA}, 1400~\text{\AA}, and 1425~\text{\AA} indices remain comparatively stable. The 1719~\text{\AA} and 1853~\text{\AA} features also show limited variation, although their behaviour depends on the detailed balance of transitions contributing to each blended index. Nebular emission generally acts as a secondary effect, but the 1460~\text{\AA} and 1853~\text{\AA} indices are significantly affected by changes to the local continuum and by blending within the index and pseudo-continuum windows.

The \flares\ analysis shows that the metallicity definition adopted for simulated galaxies is important when interpreting UV equivalent widths. Total mass-weighted metallicity traces the integrated enrichment history of the galaxy, while young-star and UV-luminosity-weighted metallicities emphasise the populations most directly connected to the emergent UV spectrum. The resulting mass--metallicity relations follow the expected trend in which more massive systems are more metal enriched, while lower-mass, high-sSFR galaxies remain more metal poor and more strongly affected by recent star formation.

Overall, the 1719~\text{\AA} index emerges as one of the most stable UV metallicity tracers across metallicity, star formation history, nebular emission, and redshift. The 1370~\text{\AA}, 1400~\text{\AA}, and 1425~\text{\AA} indices also perform robustly, although the 1425~\text{\AA} feature exhibits mild high-metallicity non-linearity. The 1460~\text{\AA} index provides strong sensitivity but is highly susceptible to burst-driven and nebular effects. The 1501~\text{\AA} and 1550~\text{\AA} indices retain diagnostic value but show increased variance due to their sensitivity to recent star formation and stellar-wind-related effects. The 1533~\text{\AA} and 1853~\text{\AA} features remain useful in enriched systems, although the 1853~\text{\AA} index requires caution where nebular contamination or line blending is significant.

At present, direct measurements of stellar metallicity in the reionization era remain observationally challenging, and constraints are often inferred indirectly or through forward-modelled synthetic spectra. The rapid expansion of rest-frame ultraviolet spectroscopy from recent \textit{JWST} surveys at $5 \lesssim z \lesssim 10$ provides an opportunity to test these physically motivated metallicity diagnostics against observed high-redshift galaxy spectra. Applying the framework developed here to \textit{JWST}/NIRSpec samples will enable direct comparison between simulated predictions and observed UV absorption features.

We show that rest-frame UV absorption indices provide a promising set of stellar metallicity tracers when interpreted within a forward-modelled framework. The use of multiple indices is particularly important, as it mitigates the limitations of any single diagnostic and allows more robust metallicity inference in galaxies with complex star formation and enrichment histories. While individual features may be weakened, blended, or affected by nebular emission, the combined behaviour of the index suite preserves clear information about stellar metallicity.

A limitation of the present investigation is the absence of explicit $\alpha$-enhancement in the simple stellar population analysis of Section~\ref{sec:theory}. Variations in $[\alpha/\mathrm{Fe}]$ provide an important probe of chemical enrichment histories, particularly in high-redshift galaxies where rapid star formation can produce non-solar abundance patterns. Future work will extend this analysis by examining the behaviour of UV indices under varying $\alpha$-enhancement in both idealised stellar populations and hydrodynamical simulations. This will allow a more detailed assessment of how abundance variations affect the interpretation of UV absorption features in early galaxies.

The work done provides a basis for applying UV indices to \textit{JWST}/NIRSpec observations in order to constrain stellar metallicities in high-redshift galaxies. Such an application requires sufficiently large samples and adequate signal-to-noise ratios, however, offers a direct route for testing whether the trends predicted by \texttt{Synthesizer} and \flares\ are present in observed spectra. Upcoming analyses will therefore focus on comparing noise-modelled cosmological simulations with high-redshift \textit{JWST}/NIRSpec observations, testing UV indices as probes of chemical enrichment in the early Universe.

\section*{Acknowledgements}
We thank the \eagle\, team for their efforts in developing the \eagle\, simulation code.

This work used the DiRAC@Durham facility managed by the Institute for Computational Cosmology on behalf of the STFC DiRAC HPC Facility (www.dirac.ac.uk). The equipment was funded by BEIS capital funding via STFC capital grants ST/P002293/1, ST/R002371/1 and ST/S002502/1, Durham University and STFC operations grant ST/R000832/1. DiRAC is part of the National e-Infrastructure. CMB thanks STFC for support through STY/X001121/1. APV, WJR, and SMW acknowledge support from the Sussex Astronomy Centre STFC Consolidated Grant (ST/X001040/1). 

We wish to acknowledge Christopher Lovell, Elizabeth Stanway, Louise T. C. Seeyave, and James Trussler for their advice and support. 

We also wish to acknowledge the following open source software packages used in the analysis: \textsc{Scipy} \cite[][]{2020SciPy-NMeth}, \textsc{Astropy} \cite[][]{robitaille_astropy:_2013}, and \textsc{Matplotlib} \cite[][]{Hunter:2007}

We list here the roles and contributions of the authors according to the Contributor Roles Taxonomy (CRediT)\footnote{\url{https://credit.niso.org/}}.
\textbf{Connor Sant Fournier, Joseph Caruana, Stephen M. Wilkins, Aswin P. Vijayan}: Conceptualization, Data curation, Methodology, Investigation, Formal Analysis, Visualization.
\textbf{Kristian Zarb Adami, Conor Byrne, Jack C. Turner, William J. Roper, Aswin P. Vijayan}: Writing - review \& editing.

\section*{Data Availability}

The data associated with the paper is publicly available at \href{https://flaresimulations.github.io}{https://flaresimulations.github.io} and \href{https://synthesizer-project.github.io/synthesizer/index.html}{https://synthesizer-project.github.io}.



\bibliographystyle{mnras}
\bibliography{flares, flares-uvindices} 







\label{lastpage}
\end{document}